\documentclass[%
 reprint,
 article,
 amsmath,amssymb,
 aps,
]{revtex4-1}

\usepackage{graphicx}
\usepackage{dcolumn}
\usepackage{bm}
\usepackage{color}

\usepackage{soul}

\begin{document}


\title{Nanoscale Liquid Crystal Lubrication Controlled by Surface Structure and Film Composition}

\author{Pritam Kumar Jana$^{1}$}
 \email{pritam.jana@aalto.fi}
 
\author{Wei Chen$^{2}$}
\author{Mikko J. Alava$^1$}%


\author{Lasse Laurson$^{1}$}
 \email{lasse.laurson@aalto.fi}

\affiliation{$^1$COMP Centre of Excellence, Department of Applied Physics, Aalto University, P.O.Box 11100, FI-00076 Aalto, Espoo, Finland}
\affiliation{$^2$State Key Laboratory of Multiphase Complex Systems, Institute of Process Engineering (IPE), Chinese Academy of Sciences (CAS), Beijing 100190, PR China.}

\begin{abstract}



Liquid crystals have emerged as potential candidates for next-generation 
lubricants due to their tendency to exhibit long-range ordering. Here, we construct a full 
atomistic model of 4-cyano-4-hexylbiphenyl (6CB) nematic liquid crystal lubricants mixed 
with hexane and confined by mica surfaces. We explore the effect of the surface 
structure of mica, as well as lubricant composition and thickness, on the nanoscale friction
in the system. Our results demonstrate the key role of the structure of the mica 
surfaces, {\color{black}specifically the positions of potassium ($\mathrm{K}^+$)} ions, in determining the nature of sliding friction with monolayer lubricants, including 
the presence or absence of stick-slip dynamics. {\color{black} With the commensurate setup of confining surfaces, when the grooves created between the periodic $\mathrm{K}^+$ ions are parallel to the sliding direction we observe a lower friction force  as compared to the perpendicular situation. Random positions of ions exhibit even smaller friction forces with respect to the previous two cases.} 
For thicker lubrication layers the 
surface structure becomes less important 
{\color{black} and we observe a good agreement with the experimental data on bulk viscosity of 6CB and the additive hexane. In case of thicker lubrication layers, }
friction may still be controlled by tuning the 
relative concentrations of 6CB and hexane in the mixture. 

\end{abstract}

\pacs{Valid PACS appear here}
\maketitle


\section{Introduction}
Controlling friction, wear, and lubrication by understanding the atomic-scale processes taking 
place at the interfaces of interacting bodies in relative motion has been a long-standing 
challenge, with applications, for example, in micro- and nanoelectromechanical 
systems \cite{Bhushan, Park, Bhushan1}.   
Many classical laws of friction and lubrication 
in such systems are violated due to the high surface-to-volume ratio and the greater importance 
of molecular interactions and arrangements in determining the surface forces \cite{Smith}. 
As a result, a number of fundamental questions in this field are still unsolved \cite{Urbakh}.

A practical design goal for applications is to reduce stiction, friction and wear. 
To this end, one may think of two main strategies: either modulating the roughness, electrostatic interactions, 
crystal structure, edge orientations and other properties of the surfaces that come into contact or by using 
various types of lubricants between the surfaces\cite{Frenken,C7NR07839K}. 
Approaches with the idea of tuning the potential energy landscape between the interacting surfaces at 
the atomic scale have been proposed \cite{Frenken}, including the use of aperiodic 
quasicrystal surfaces or introducing a lattice mismatch between the two sliding crystal 
surfaces, leading to the superlubricity mechanism \cite{Park, Filippov, Hirano}. However, a recent study shows that the duration of the superlubric state, i.e., the incommensurate configuration can be finite and therefore, ultralow friction does not prevail. The possibility of rotation of the sliding surface stabilizes the high frictional commensurate configuration \cite{FilippovPhysRevLett.,WoodsNatComm}. The idea of aligning the crystallographic orientation of confining layers is used in resonant tunneling diodes \cite{Mishchenko} and other novel devices \cite{WoodsNatComm}. 
It suggests to perform more investigations on reducing friction considering energetically stabilized high frictional commensurate setup. Guo et al. have shown that interlayer friction, in case of commensurate setup, can be reduced by functionalizing the sliding surfaces \cite{Guo}.
Several studies have been performed focusing on the quest of perfect lubricants and their dynamical properties or the adsorbate surface coverage while being sheared by the confining surfaces in relative sliding motion \cite{Bhushan,Thompson1,zaidan2016mixture,C7NR09530A}.

Liquid crystals (LCs) have been explored as potential lubricant due to ultra-low friction 
as a result of their long-range orientational ordering tendency \cite{Amann}. 
Due to the high cost of pure LCs, various mixtures of LCs with other substances
are typically considered, and more extensive investigations are required to understand the 
effects due to such additives \cite{Manzato, Strelcov, Hsuan_Wei, Nakano, Chen}. 
There are open questions related to details of the interactions of the LC 
molecules with the confining surfaces, and the phase behavior they exhibit under
applied shear. While experiments, coarse-grained simulations, and theoretical studies to understand the 
structural and dynamical properties of LC lubricants have been performed 
\cite{BERMUDEZ_1997, IGLESIAS_2004, Carrion_2009, Manzato, Ruths, Noirez, Richard, Robic, Idziak1915, Cheng10062008, Zhang_2015, Chen}, 
atomistic simulations to establish a proper link between coarse-grained models and experiments are
still missing.  

In this article, we focus on the lubricating properties of nematic 4-cyano-4-hexylbiphenyl (6CB) LCs in pure and in presence of hexane additives where  mica serves as confining surfaces by using fully atomistic model simulations. 
We probe the influence of the relative orientation of the surfaces including
an incommensurate setup, and the effect of the random ion distribution in the confining mica surfaces.  We also inspect the impact of the thickness of the lubricant film, whether pure 6CB or a 6CB/hexane mixture, on frictional response. 
Our results show that friction in systems with monolayer lubricant films is sensitive to the arrangement of ions on the confining 
surfaces or to the relative surface orientations. Moreover, we demonstrate friction control also for thicker lubricant layers via tuning the composition of the LC-hexane mixture, including controlling the presence of stick-slip.
\section{Model}
{\color{black}There are several efforts to reproduce the properties, for example, the so-called odd-even effect in nematic-isotropic transition, of liquid crystal homologues namely 4-cyano-4-alkylbiphenyl (nCB) up to a satisfactory level of accuracy in computer simulations\cite{cifelli2008atomistic,tiberio2009towards}. In the present study, we are specifically interested in the lubricating properties of liquid crystals because of its long-range ordering. To this end, we choose 4-cyano-4-hexylbiphenyl (6CB). There is experimental evidence on lubrication of 6CB confined between mica surfaces\cite{janik1997shear,kitaev2000thin,mizukami2004shear}. We could have chosen other homologues. However, we believe that the observed results will qualitatively remain same. We consider hexane as an additive because of its low cost and is largely used as a lubricant.}
To construct the atomistic models of 6CB and hexane molecules, we use force field 
parameters from Refs. \cite{Adam,Cheung}. As for the confining mica surfaces, we consider 
2M1-muscovite mica with the formula KAl$_2$(Al,Si$_3$)O$_{10}$(OH)$_2$, using the force 
field parameters as described in Ref. \cite{Heinz}. {\color{black}Details of the force field parameters used in the simulations are provided in the Supporting Information. To verify the accuracy of the force field, we have calculated two friction related quantities: (a) bulk viscosity for both 6CB liquid crystal and hexane and we see a good agreement with experiments. See the inset of Fig. \ref{Fig:Confined-bulk}. (b) friction force for commensurate (confining surfaces are aligned) and the incommensurate surface (confining surfaces are misaligned) and we observe a lower friction in case of the incommensurate surface which is also observed in experiments. Both results, we discuss in detail in Results and discussions section. It assures that the used force field parameters are good enough to study frictional properties of 6CB and hexane. } 

\begin{figure}[t!]
\centering 
 \includegraphics[ scale=0.375]{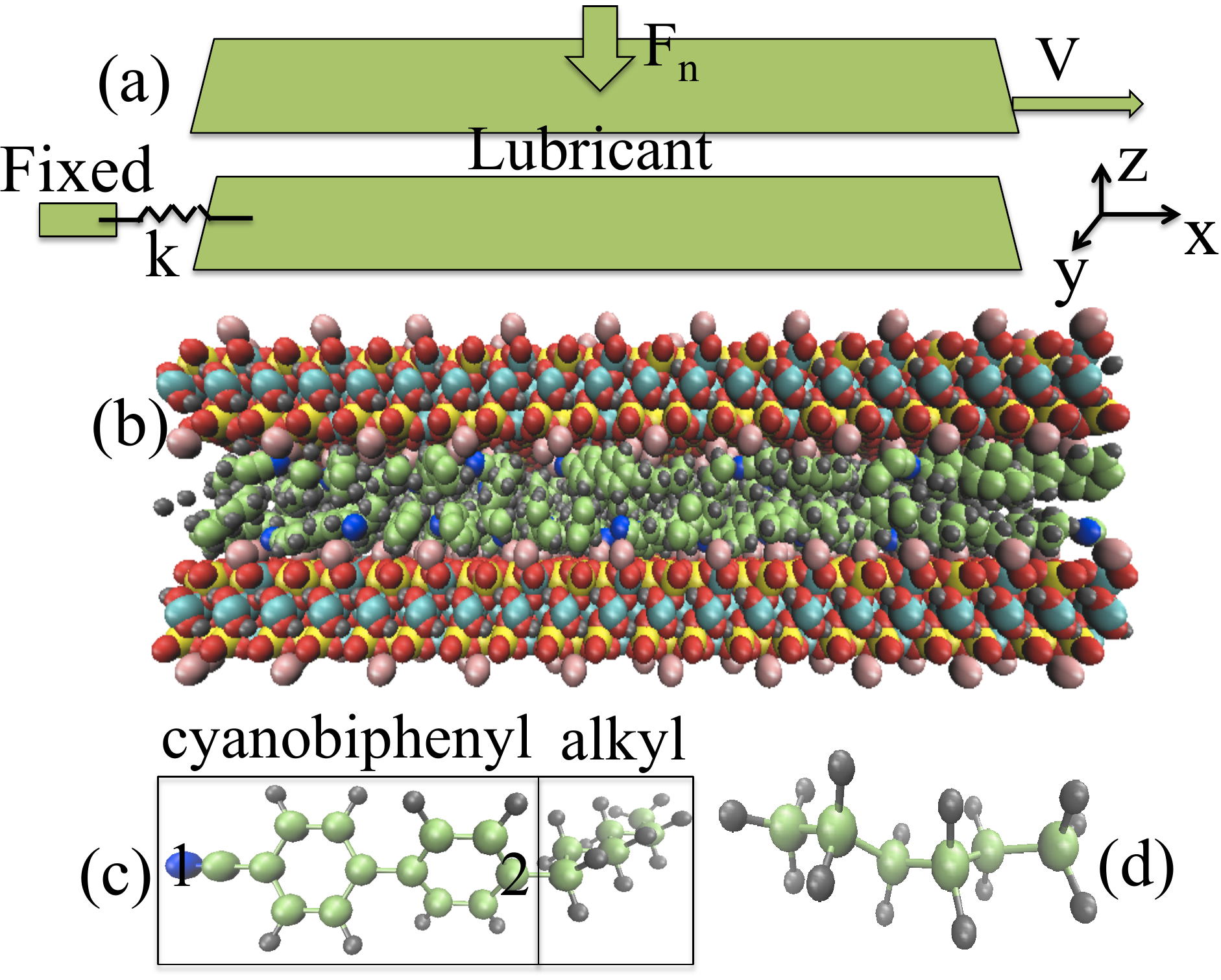}  
 \includegraphics[ scale=0.275]{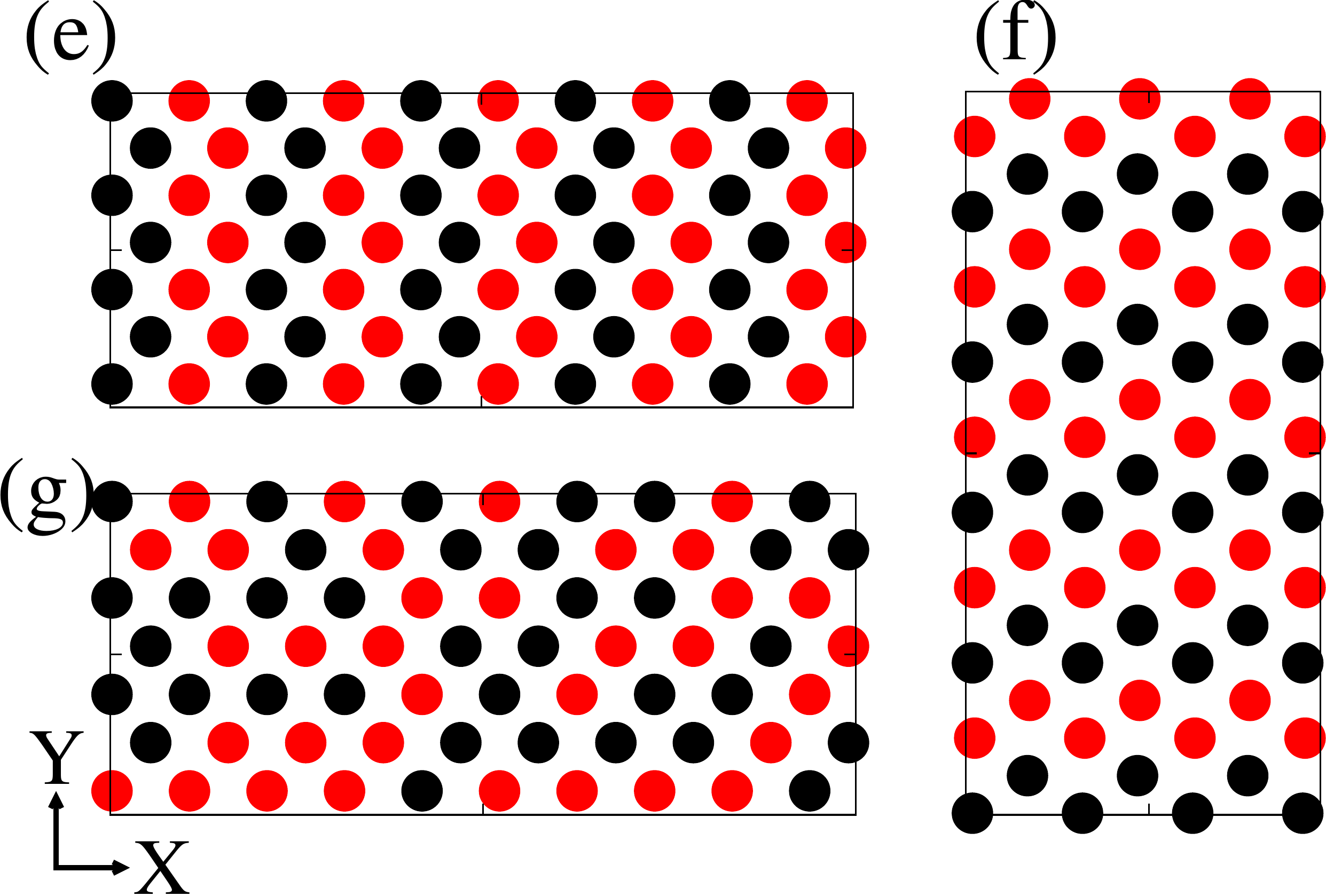}
\caption {(a) A schematic of the model. Lower plate is attached with a spring to a fixed 
point. The upper surface is subject to a normal load $F_{\text{n}}$ and driven along the 
$+x$ direction with a constant velocity $V$. (b) A snapshot of the system is depicted, 
with $144$ 6CB molecules confined by the two mica plates. (c) and (d) correspond 
to the molecular structures of 6CB (consisting of cyanobiphenyl and alkyl groups) and hexane, 
respectively. Pink, red, yellow, cyan, green, grey and 
blue colored atoms are potassium, oxygen, silicon, aluminum, carbon, hydrogen and 
nitrogen, respectively. (e), (f) and (g) show the different arrangements of the 
K$^+$ ions at the surfaces considered in the simulations (only a small part of the surface
shown, with red and black dots corresponding to upper and lower faces of the mica sheet, 
respectively), i.e., with grooves perpendicular and parallel to the sliding direction, 
and a random arrangement of the ions, respectively.}
 \label{Fig:Model}
\end{figure}

{\color{black} In an experiment, Atomic Force Microscope (AFM) cannot detect potassium ions (K$^+$) on mica surfaces. A probable reason of that  is K$^+$ ions  on mica are moved around by the AFM cantilever-tip during the sliding motion\cite{xu1998effects}.  However, there are several efforts on understanding the spatial arrangement of ions at the solid-liquid interface. Ricci et al. have shown that the monovalent metal ions do not adsorb on mica surfaces immersed in water randomly but form preferential ordered structures to minimize the surface energy at the mica-ion interface\cite{ricci2014water}. Most of the theoretical studies consider the periodic arrangement of K$^+$ ions\cite{Odelius}. However, due to the randomness of the cleaving process of mica, uniform distribution of K$^+$ ions is unlikely\cite{bampoulis2017graphene}. Those experimental observations lead us to consider both periodic and random arrangements of K$^+$ ions and examine the resulting effects on friction.}
Each surface consists of $10\times 6$ 
unit cells and linear dimensions of $L_x= 103.76$ \AA \ and $L_y=70.54$ \AA, respectively. 
Both surfaces are kept parallel 
to the $xy$ plane, and the upper surface is driven with a constant velocity of $V = 0.1$ 
m/s along the $x$ direction. We do not consider the $y$ directional motion of any of the plates. As shown in Fig. \ref{Fig:Model}(a),
the lower surface is attached to a fixed point with a spring constant 
$k/N_{\text{p}}=0.007$ N/m, with $N_{\text{p}}$ the number of atoms in the surface which is a basic tribological setup \cite {Ouyang}. The spring constant is consistent with experimental values as mentioned in Refs. \cite {Dienwiebel, Yoshizawa} and the sliding velocity (here 0.1m/s) of the upper plate is smaller than the critical velocity defined as the limiting velocity after which  stick-slip motion disappears \cite{Lei}. Therefore, under the chosen values of  parameters, we expect stick-slip dynamics in the system with thin layer of lubricants so that in addition to study the frictional effect of ions position on mica surfaces, we can also characterise the dynamical behavior of nano-confined liquid crystals during the stick and the slip events.

In the steady state, the upper surface is subject to a normal 
load corresponding to a 1 atm. {\color{black} We do not vary normal load. However, it could be an interesting direction to explore the responses of 
nanoscale friction for different normal loads that usually disobey a simple linear relation\cite{Urbakh}.}Temperature of the lubricants is maintained at $T=298$ K using a Langevin thermostat, applied only in the $y$ direction to avoid streaming bias 
\cite{Thompson1,thompson1990shear,chen2015stick}. 
To make sure that thermostat does not affect our main results, we use the temperature relaxation time $10^{-4}$ ns which is significantly smaller than the time required ($\sim 10^{-2}$ ns) to finish a slip event. 
The equations of motion are solved with the velocity Verlet algorithm implemented in the 
LAMMPS code\cite{plimpton1995fast}, with an integration time step of 1 fs. Long-range electrostatic interactions 
are computed by using the particle-particle-particle-mesh solver for the slab geometry \cite{patel2006molecular,pan2017effect} with $10^{-4}$ accuracy {\color{black}\cite{cerda2009understanding} implemented in LAMMPS\cite{plimpton1995fast}}. 
Initially lubricant molecules are arranged in a simple cubic lattice and first 
equilibrate for 100 ps with both confining surfaces fixed. During the following 100 ps, the 
top surface is subject to a normal load corresponding to a 1 atm pressure and is allowed 
to move along the $z$ direction. After that, the top surface is driven along the $x$ axis 
with a velocity $V$ for 60 ns; in most cases the observables of interest
have been averaged over the last 20 ns of 
the simulations, to ensure that a steady state is reached.

\section{Results and discussions}
{\color{black}We divide our results into two subsections. First, we discuss the effects of surface structure and orientation on friction. Next, we show how  additives' concentration can tune the resulting friction forces. }
\subsection{Effects of surface structure and orientation on friction}
{\color{black} As we are interested in understanding the boundary lubrication, where friction between the confining surfaces depends on the surface properties and the properties of the thin layer of lubricants, first we consider the system with monolayer of 6CB and hexane molecules.}
For the system geometry considered here, monolayer 
systems are obtained by considering 72 6CB or 144 hexane molecules, respectively. To construct 
a monolayer we choose the number of molecules to be such that the total area occupied by them 
is smaller than the area of the confining surfaces. 
To characterize the dynamics, we compute the friction force $F_{\text{s}}$ (the average force 
exerted on each atom of the bottom plate by the lubricant and the top mica plate along the 
$+x$ direction), film thickness $D$ 
(distance between the two plates), and the order parameter $S$ of the LCs, i.e., the 
maximum eigenvalue of the average ordering tensor $Q_{\alpha,\beta}$ defined as  
\begin{equation}
Q_{\alpha,\beta}=\frac{1}{N}\sum_i^N\left(\frac{3}{2}u_\alpha^i u_\beta^i-\frac{1}{2}\delta_{\alpha,\beta}\right),
\end{equation}
where $u_\alpha^i ({\alpha,\beta}=x,y,z)$ are the Cartesian components of the unit vector of 
the LC molecule $i$, $N$ is the number of LC molecules, and $\delta_{\alpha,\beta}$ is the 
Kronecker delta. Here, the unit vector of the LC molecule is taken to be parallel to the 
line connecting the nitrogen atom and the end carbon of the cyanobiphenyl group of 6CB, 
denoted by 1 and 2 in Fig. \ref{Fig:Model}(c), respectively. We also compute the 
director (eigenvector of the maximum eigenvalue) of the LC molecules (the average molecular 
orientation), and characterize it by the angle $\phi$ with the $x$-axis.

\begin{figure}[t!]
 \centering
  \includegraphics[scale=0.3]{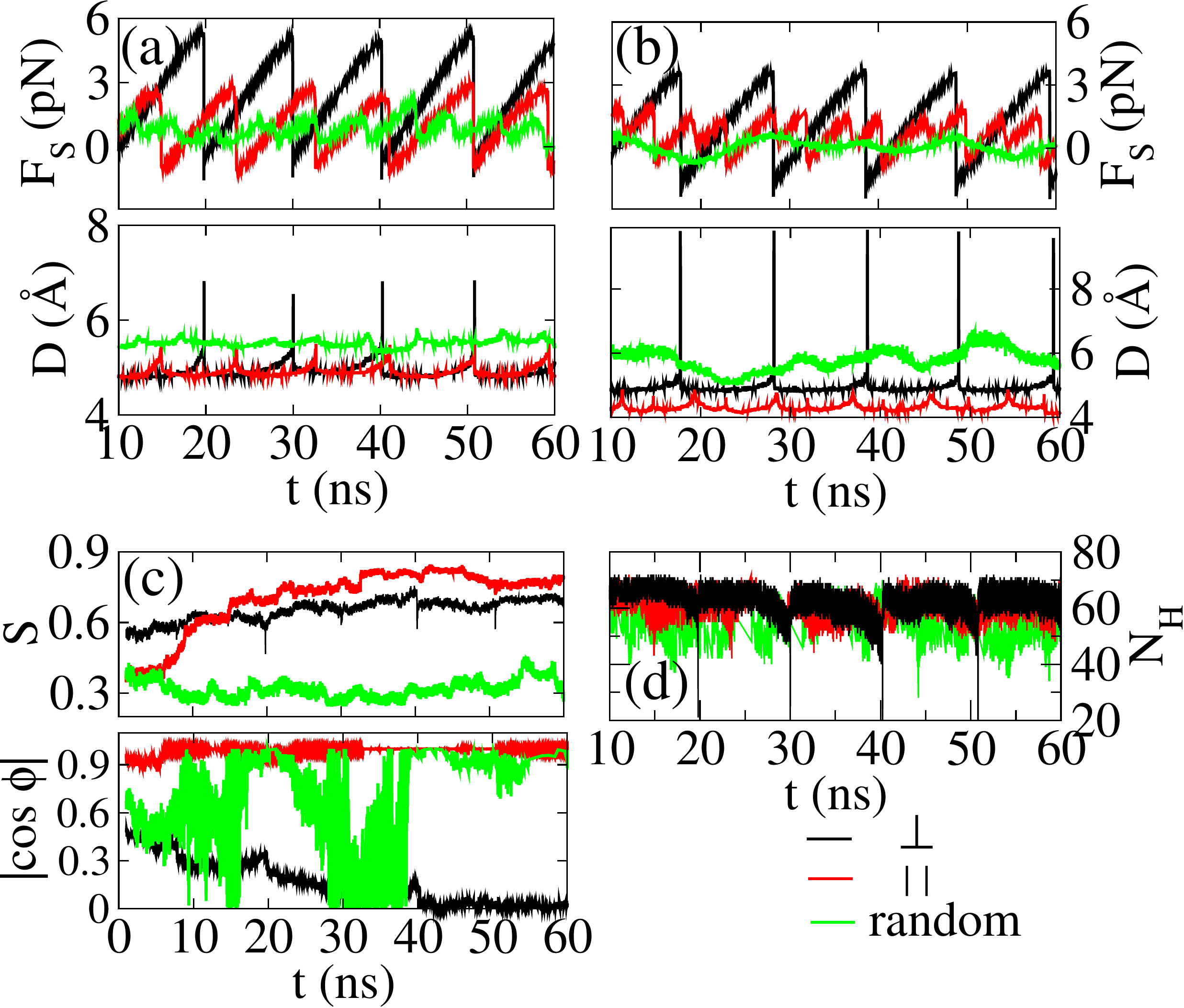} 
\caption {Time evolution of the friction force $F_{\text{s}}$ and film thickness $D$ 
for monolayers of (a) 6CB and (b)
hexane, considering the three different arrangements of the K$^+$ ions at the surfaces depicted 
in Fig. \ref{Fig:Model}. (c) Order parameter $S$ and the angle $\phi$ between the 6CB 
director and the sliding direction, quantified here by $\cos \phi$, are shown in the 
top and the bottom panel, respectively. (d) Time evolution of the number 
$N_{\text{H}}$ of hydrogen bonds between the 6CB and bottom mica, illustrating that the 
bonds tend to break during the slip events.}
\label{Fig:Surfaceeffect}
\end{figure}

The monolayer results are summarized in Fig. \ref{Fig:Surfaceeffect}, 
considering three different arrangements of the K$^+$ ions on the mica surfaces: Figs. 
\ref{Fig:Model}(e) and (f) display two different periodic arrangements of the 
ions [the structure in (f) is obtained by rotating the surface in (e) by 90$^{\circ}$], 
while Fig. \ref{Fig:Model}(g) shows an example with randomly positioned ions. 
It is important to notice that in the case of the two periodic arrangements, grooves are 
generated along the $x$ [Fig. \ref{Fig:Model}(f)] or $y$ direction [Fig. 
\ref{Fig:Model}(e)] in between stripes of ions, resulting in an anisotropic 
surface structure, while in the case of randomly positioned ions, such features are absent 
[Fig. \ref{Fig:Model}(g)]. To examine the effect of surface structure on 
boundary lubricated friction, we plot in Figs. \ref{Fig:Surfaceeffect} (a) and (b) the 
time-dependence of the friction force $F_{\text{s}}$ and the film thickness $D$ for those 
three different structures of the confining mica surfaces for 6CB and hexane, respectively; 
see also example movies (Videos SM1, SM2, and SM3) provided as Supporting Information. We observe regular 
stick-slip dynamics for both 6CB and hexane for the ordered surfaces. The maximum 
friction force (and thus the magnitude of the stick-slip oscillations) as well as the maximum 
film thickness during the ``jumps'' of the top plate associated with the slip events are 
larger in the case where the grooves of the mica surfaces are perpendicular to the 
sliding direction. In case of 6CB lubricated system, this effect is visible also when considering the time evolution of
the number of hydrogen bonds, $N_{\text{H}}$ , formed between 6CB
hydrogens and the bottom mica plate (i.e., the number of 6CB hydrogens closer 
than 3 \AA \ from the bottom mica surface \cite{chen2015stick}. A typical hydrogen bond length is ~1.5  \AA \  to 2.5  \AA \ which is smaller than the distance, 3 \AA \ , that we consider as the breakage of hydrogen bonds.), see Fig. \ref{Fig:Surfaceeffect} 
(d): bonds break as the system evolves from stick to the slip state.  
In the case of the randomly positioned ions, 6CB exhibits more irregular stick-slip dynamics, 
with also some visible jumps in $D$ accompanying the slip events, while no clear signature of 
stick-slip is observed for hexane. Due to the incommensurate nature of the confining 
surfaces with randomly positioned K$^+$ ions, the average film thickness $D$ is larger 
and the maximum $F_{\text{s}}$ is lower than with ordered mica surfaces. {\color{black} A similar observation is depicted in an experiment where surface ion induced tribological properties have been studied. It has been shown that when the ions are strongly bound and randomly distributed on mica irregular stick-slip occurs \cite{xu1998effects}. Several experimental studies have been performed to understand the dynamical or mechanical properties of  organic liquids when they are confined to few layers by solid surfaces such as mica \cite{kumacheva1998simple}. A common phenomenon is the observation of stick-slip motion depending on the sliding velocity and the normal load \cite{ohnishi2007influence, rosenhek2015question}.  Although there are several efforts, molecular origin of stick-slip cycle in sheared solid-like lubricants is not well understood because of the difficulties 
to capture the behaviour of lubricants during the slip events, precisely  mainly for two reasons:
(a) slip events occur in nanometrically confined film (b) slip events are of very short duration 
and occupy a tiny fraction of the stick-slip cycle \cite{rosenhek2015question}. Therefore, a lot of understanding has been derived from theoretical and computer simulations. To this end, we also explore the behaviour of confined LCs during the stick and the slip events.}

For 6CB LCs, the time-evolution of the order parameter $S$ and the angle $\phi$ between 
the LC director and the sliding direction ($x$ axis) [Fig. \ref{Fig:Surfaceeffect} (c)] 
encode additional information about the dynamics. In particular, 6CBs have a tendency to 
orient along both the grooves of the confining surfaces (due to formation of hydrogen 
bonds between the biphenyl hydrogen and the surface oxygen exposed along the grooves), as well as 
along the sliding direction: $S$ is larger and $\phi$ is smaller when the grooves and the 
sliding direction are parallel [both along $x$, see Fig. \ref{Fig:Model}(f)]. 
For the perpendicular case, in the steady state the 6CBs point mostly along $y$ (i.e., along the grooves), 
but $S$ is a little smaller than in the parallel case due to the competing ordering mechanism 
of the sliding along $x$. 
{\color{black} It shows that the monolayer of LCs prefers to orient along the microgrooves instead of aligning along the sliding direction due to the presence of strong chemical interactions.}
In contrast, for the random arrangement of K$^+$ ions on 
mica, the 6CBs exhibit clearly less orientational order (smaller $S$), and $\phi$ also 
fluctuates significantly in time, see Fig. \ref{Fig:Surfaceeffect} (c). 
{\color{black} The alignment of LCs along the grooves is not a complete surprise. Vegt et al. have investigated the orientation of 
4-cyano- 4-octylbiphenyl (8CB) liquid crystals on anisotropic polyimide surfaces by performing molecular dynamics simulations where they have shown that a single molecule of 8CB prefers to orient along the microgrooves because of the strong binding between polar cyano groups from LCs and the carbonyl groups from the polyimide surface\cite{van2001orientation}. In general, understanding the orientation of LCs on the surface is interesting because of its application in liquid crystal display.}

{\color{black}To explore more about the dynamics of LCs and their response during the stick-slip cycle we measure the mean square displacement (MSD) of 6CB molecules.} It shows that while the 
molecules move ballistically along $x$ for time scales longer than the stick-slip period
for all mica surface structures 
(Fig. SM2 (a) in the Supporting Information) 
they exhibit diffusive dynamics in the $y$ direction in the long-time limit 
only for random ion arrangement and grooves perpendicular to the sliding direction 
(Fig. SM2 (c) in the Supporting Information). 
In contrast, for grooves parallel to the sliding direction, 
MSD saturates to a value comparable to the groove spacing, suggesting that in that case 
the grooves act as barriers for the $y$-directional diffusion of 6CBs. 
However, the slip is confined to the lubricant close to the lower plate, and happens in the direction of sliding. Details are shown in the Supporting Information. 
Thus, for monolayer lubricants, the structure of the confining surfaces plays a decisive role in the 
frictional response. So far, we have discussed the system where confining surfaces are aligned with each other- thus commensurate. When we consider the incommensurate setup, i.e, two surfaces are misaligned with respect to each other, we see a frictional response which is even somewhat lower than the situation where K+ ions are randomly positioned on the surface. See Fig. SM7 in the Supporting Information. {\color{black}Experiments show the similar observations as misaligned mica surfaces exhibit lower friction forces as compared to the commensurate setup \cite{Filippov, Hirano}}. 

\begin{figure}[t!]
 \centering
   \includegraphics[scale=0.4]{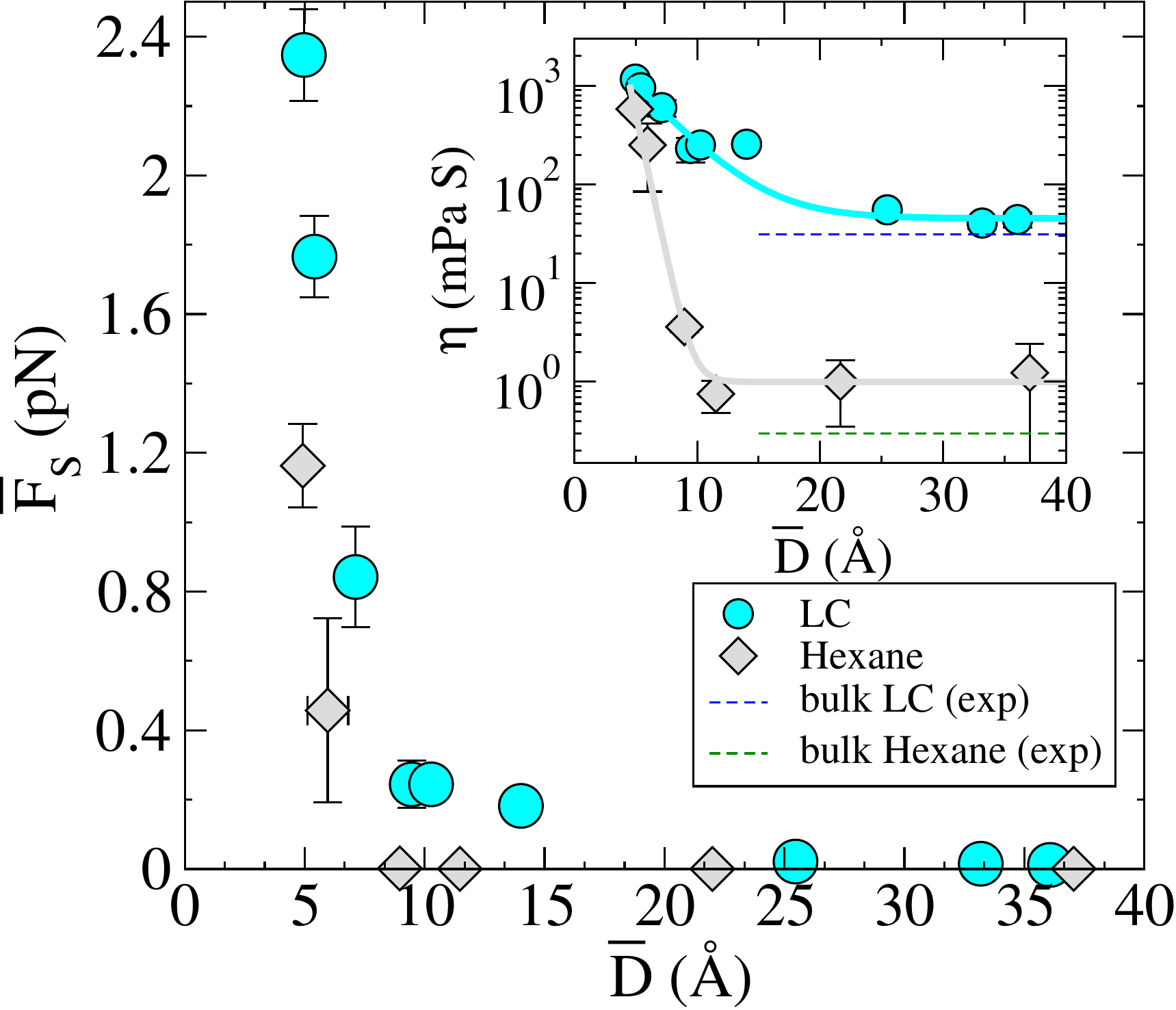} 
\caption {Average friction force $\bar F_{\text{s}}$ as a function of the average 
film thickness $\bar D$, obtained by considering different numbers of molecules 
confined by the mica surfaces, with the film consisting of pure 6CB or 
pure hexane. The inset shows the corresponding dynamic viscosities 
$\eta$, which approach in both cases the experimental bulk viscosity values 
(Refs. \cite{Hirano2,Jan} and \cite{rodriguez2006dynamic} for 6CB and hexane, 
respectively), indicated by the dashed horizontal lines. Solid lines are fits of 
the form of $\eta (\bar D) \propto \exp (-\bar D / \lambda) + \eta_0$, yielding 
$\lambda = 0.73$ \AA \ and 3.36 \AA \ for hexane and 6CB, respectively.}
 \label{Fig:Confined-bulk}
\end{figure}

To understand the effect of the confining films' thickness on friction, we study 
here the periodic pattern of the K$^+$ ions with the grooves perpendicular to the sliding 
direction; for thicker lubricant films, stick-slip dynamics gradually disappears with
increasing $D$ (see Fig. SM3 in the Supporting Information), and the surface structure of mica becomes less important, 
with all the three surface structures from above yielding similar 
results for the friction force. {\color{black} A similar observation is reported when water is confined between mica surfaces \cite{chen2015stick}.} 
We consider various systems with the number of 6CBs 
ranging from 72 to 600, and the number of hexane molecules from 144 to 640, and also 
systems with a smaller area of the confining mica plates to reach a larger 
$D$ for a given number of lubricant molecules. We compute the resulting average friction force 
$\bar F_{\text{s}}$ as a function of the average thickness $\bar D$ (Fig. \ref{Fig:Confined-bulk}; 
see also Supporting Information Video SM4 for an example movie of a thick 6CB system). 
The inset of Fig. \ref{Fig:Confined-bulk} displays the corresponding dynamic viscosities $\eta$, 
defined via $\bar F_s N_p=\eta A (V/\bar D)$, where $A$ is the surface area \cite{Leng}. 
\color{black}Under strong confinement (small $\bar D$), both systems exhibit a high dynamic viscosity which decreases 
with increasing $\bar D$, and approaches the known bulk viscosity values at $T$ = 298 K, 
i.e., $31.3$ and $0.3$ mPa$\cdot$s for 6CB \cite{Hirano2, Jan} and hexane 
\cite{rodriguez2006dynamic}, respectively. Notice that in general $\eta$ is expected to 
depend on $V$ (or the shear rate) \cite{jabbarzadeh2005very}, but our results indicate that 
$V=0.1$ m/s is a sufficiently low sliding velocity such that $\eta$ approaches a value close 
to that of the bulk viscosity in the limit of large $\bar D$. We fit the data by 
$\eta (\bar D) \propto \exp (-\bar D / \lambda) + \eta_0$, yielding a decay length 
$\lambda = 0.73$ \AA \ and 3.36 \AA \ for hexane and 6CB, respectively.
This slower approach to bulk behavior for 6CBs may be understood via the competition
between screening of the mica-mica interaction and the inherent ``stickiness'' of
the 6CB lubricant, with the latter manifested also as the higher bulk viscosity value
for 6CB. Due to their larger dielectric constant ($\epsilon \approx$ 9.5 for 6CB \cite{Ratna}
vs $\epsilon \approx$ 1.88 for hexane \cite{Frederick}), one would expect 6CBs to be more 
efficient at screening the electrostatic interaction between the mica plates 
\cite{lee2017scaling}. However, at the same time, the sticky nature of 6CB molecules, 
originating from the Coulomb interaction of the positively charged mica K$^+$ ions and 
the highly electronegative nitrogen atoms, as well as from the hydrogen bonding between 
phenyl hydrogen and mica oxygen atoms, hinders the rapid reduction of friction as 
$\bar D$ increases.
The difference in the screening properties of the two lubricants can also be seen by 
noticing that $\eta_{bulk}^{LC}/\eta_{bulk}^{H} \approx 35.8$, while in the monolayer case 
$\eta_{LC}/\eta_{H} \approx 2$ (with $\eta_{bulk}^{LC}$ and $\eta_{bulk}^{H}$ the 
measured $\eta$-values of LCs and hexane, respectively), suggesting (in relative terms)
a stronger surface-surface interaction through a thin layer of hexane as compared to the LC 
case. We note that a similar evolution with $\bar D$ of the viscosity of LC lubricants has 
been observed experimentally \cite{shen2001nano}. The MSD of 
the 6CB molecules shows that in thicker systems they exhibit ballistic motion
along $x$ and diffusive dynamics in the $y$ direction independent of the structure of the 
confining surfaces (Figs. SM2 (b) and (d) in the Supporting Information). 

We also investigate the molecular orientation for thicker lubrication films. 
In contrast to the monolayer case with grooves perpendicular to the sliding direction 
(see Fig. \ref{Fig:Model} (e)), we observe that the director fields for thicker LC films 
tend to orient along the sliding motion as shown in Fig. SM5 (a) in the Supporting Information. 
Due to larger distances between the confining plates, they also display a $z$-directional 
degree of freedom which is absent in the monolayer cases, see Fig. SM5 (b) in the Supporting 
Information. {\color{black}We do not see homeotropic alignment, i.e., a perpendicular arrangement with respect to the confining surfaces. Experiments also exhibit that in case of untreated mica, LCs have a tendency to be in the plane of the surface  \cite{ruths2000influence}.} 
In these studies of LCs as a lubricant we have not paid any direct attention to the general structural properties of the LC films. This concerns both eventual smectic order in thicker systems (and the corresponding dislocations) and the point defects or disclination lines in the presence of nematic order depending on the lubrication layer thickness.

\begin{figure}[ht!]
 \centering
   \includegraphics[scale=0.4]{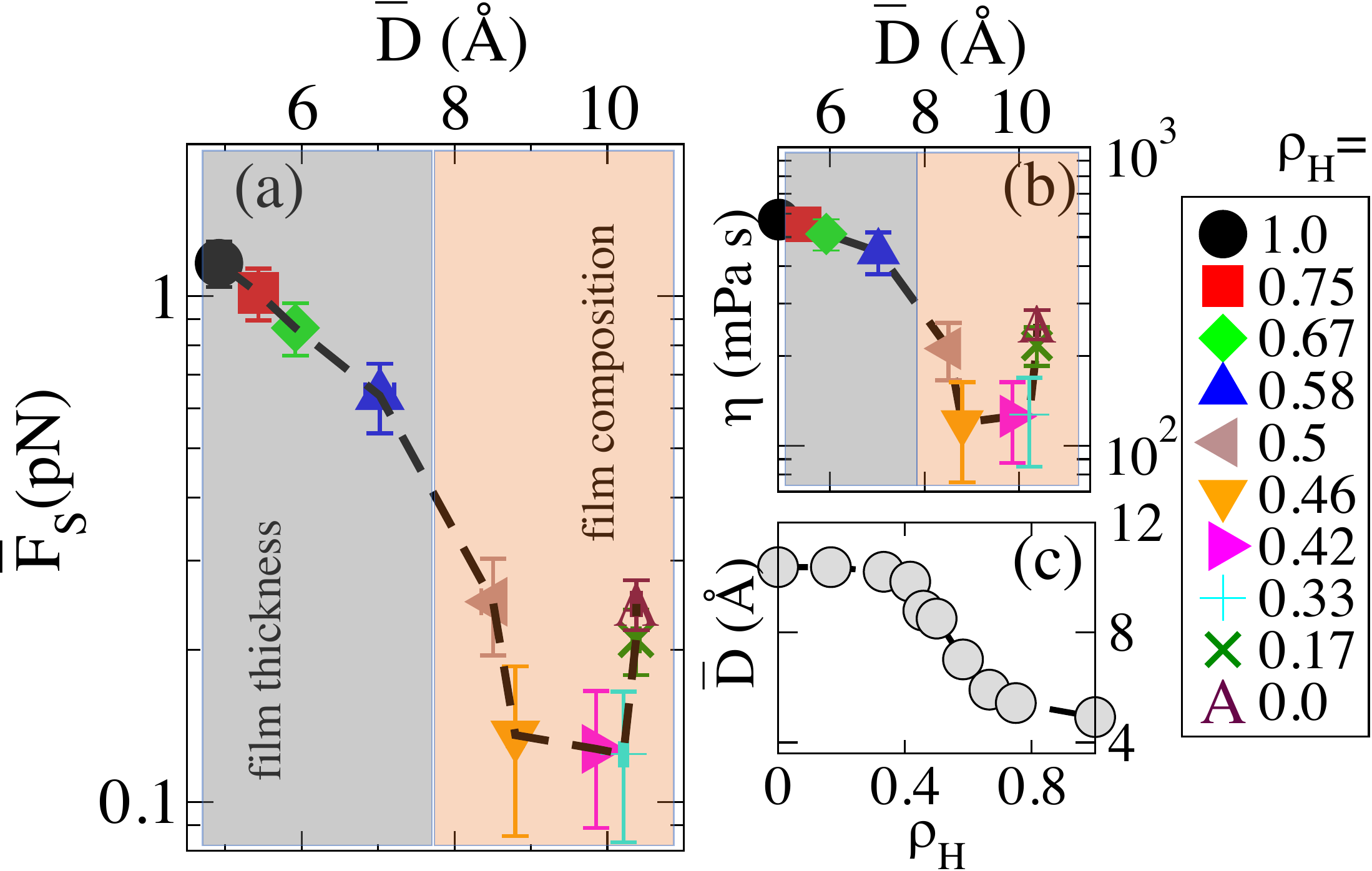} 
\caption {(a) Average friction force $\bar F_{\text{s}}$ as a function of the film thickness
$\bar D$ for films with different mixtures of 6CB and hexane. (b) shows the
corresponding dynamic viscosity $\eta$, while (c) displays the film
thickness as a function of the hexane concentration $\rho_{\text{H}}$
}
\label{Fig:Mixture}
\end{figure} 
\subsection{Effect of film composition on friction} 
Finally, to explore the potential film composition of LCs and hexane mixtures at which a reduced friction can be observed, we add various concentrations of hexane to the 
LC lubricant and compute the friction forces for different 
mixtures of 6CB LCs and hexane, see Fig. \ref{Fig:Mixture} (a). We fix the total number 
of molecules $N$ to 144, and vary the fraction $\rho_{\text{H}}=N_{\text{H}}/N$ 
of hexane from 0 to 1, where $N_{\text{H}}$ is the number of hexane molecules. 
Interestingly, the average friction force $\bar F_{\text{s}}$ displays a non-monotonic dependence 
on $\rho_{\text{H}}$ (and on $\bar D$). 
While the general trend is that $\bar F_{\text{s}}$ 
decreases with decreasing $\rho_{\text{H}}$ and increasing $\bar D$, for the smallest $\rho_{\text{H}}$ 
(corresponding to the pure and almost pure LC cases), $\bar F_{\text{s}}$ increases  
with decreasing $\rho_{\text{H}}$ and increasing $\bar D$. 
\begin{figure}[ht!]
 \centering
   \includegraphics[scale=0.4]{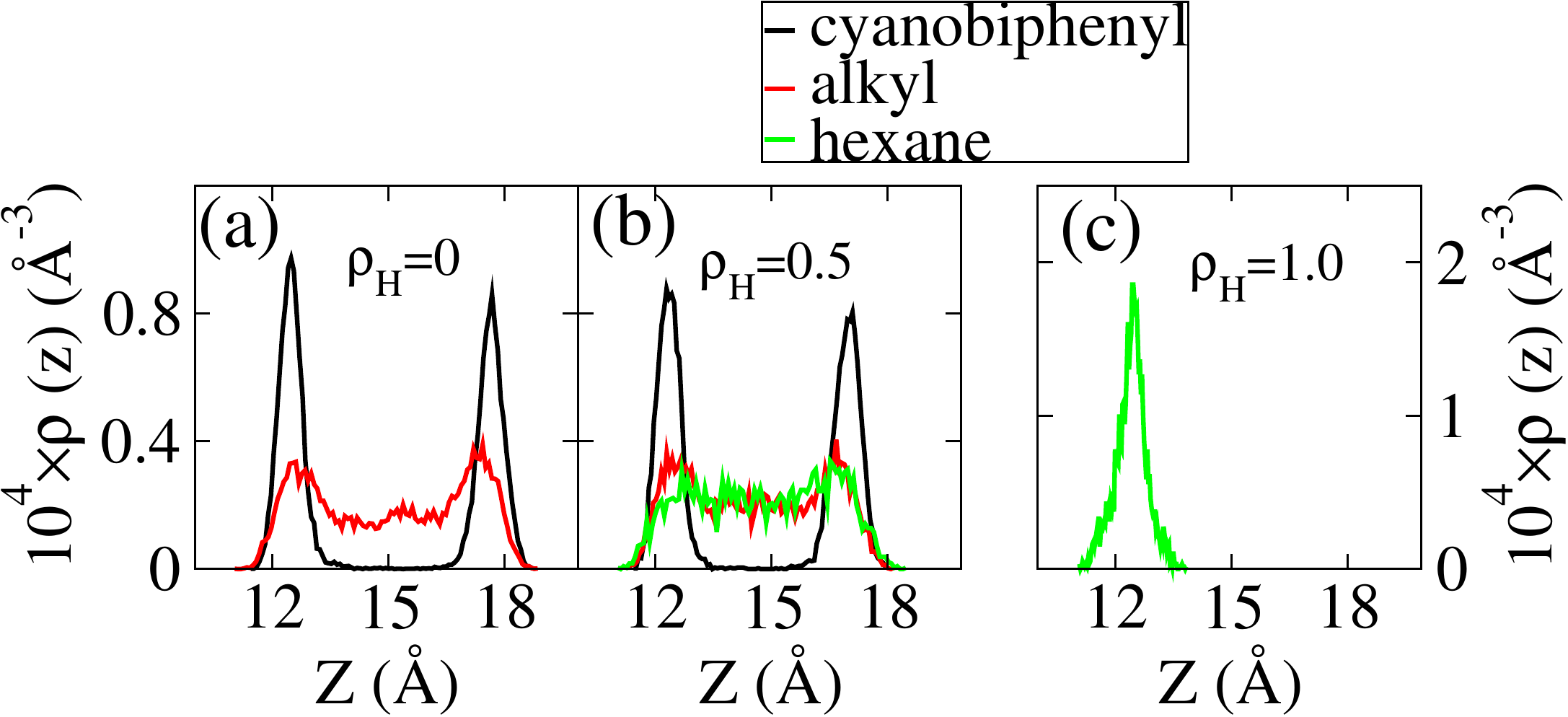} 
\caption {Probability density profiles $\rho(z)$ describing the probability per unit volume of
finding a molecule with a given $z$ coordinate, for 6CB (shown separately for 
cyanobiphenyl and alkyl groups) and hexane for three different mixtures, (a) 144 pure 6CB, (b) 72 6CB and 72 hexane mixture, and (c) 144 pure hexane. The cyanobiphenyl groups of the 6CBs tend to stay adjacent to the confining mica 
plates, while the 6CB alkyl groups as well as hexane occupy the space also in 
the middle of the gap. 
}
\label{Fig:Mixture_Probability}
\end{figure} 
{\color{black} As LCs are larger in size and stickier in nature as compared to hexane, reduction of $\rho_{\text{H}}$ or addition of LCs in the LC-hexane mixture renders a competing effect on friction. A larger size of LC will lead to the thickening of the film and thus decrease of friction while the stickier nature will oppose the effect. When $\rho_{\text{H}}=1$, hexane is confined to a monolayer between mica surfaces (see Fig. \ref{Fig:Mixture_Probability} (c)) and stick-slip motion is observed as shown in Fig. \ref{Fig:Surfaceeffect} (b), thus high friction.  
With the addition of LCs up to 1:1 ratio, film thickness increases as shown in Fig. \ref{Fig:Mixture} (c) and friction decreases rapidly (see Fig. \ref{Fig:Mixture} (a)) because of the gradual disappearance of stick-slip. See Fig. SM4 and Video SM5 in the Supporting Information.
Thus, film thickness dominates on controlling the friction. 
Note that, when $N_\text{LC}:N_\text{H}=1:1$ ($N_\text{LC}$ is the number of LCs) probability density of the confined mixtures along the $z$ direction exhibits the formation of two layers (See Fig. \ref{Fig:Mixture_Probability} (b)). Due to the strong attractive interaction between cyanobiphenyl group of 6CB and mica (K$^+$ ions and oxygens), cyanobiphenyl groups are attracted to the mica surfaces while alkyl parts as well as hexane  occupy the space in the middle in between the confining surfaces as shown in Fig. \ref{Fig:Mixture_Probability} (b).
Further addition of LCs leads to the weak changes of thickness and in that region, sticky nature of LCs plays a key role and thus a gradual increase of friction. Density profile $\rho (z)$, when $\rho_\text{H}=0$, exhibits similar probability distribution as observed when $\rho_\text{H}=0.5$. See Fig. \ref{Fig:Mixture_Probability} (a).}
The same non-monotonic behavior 
is visible in the corresponding dynamic viscosity $\eta$ [Fig. \ref{Fig:Mixture} (b)].
In precise, this non-monotonic behavior of friction is a result of the cross-over from 
a \textit{film thickness} controlled friction regime
(shown in Fig. \ref{Fig:Mixture}(a) as a gray region; 
in this regime we observe gradual disappearance of stick-slip dynamics with decreasing 
$\rho_{\text{H}}$) 
to a \textit{film composition} dependent regime (orange colored region; stick-slip not observed due to larger $\bar D$).
When we characterize the slip at surface depending on the amount of hexane, we 
see that both slip length and slip position increase as we approach the pure hexane case (See Fig. SM10 in 
the Supporting Information). The slip in confined LCs follows the surface-slip coherently. However, 
slip length at surface is larger as compared to the LCs. When the slip length at the surface is small, we do not 
see any slip in the LCs (See Fig. SM9 in the Supporting Information).

\section{Conclusions}
To summarize, we have presented an extensive study of nanoscale LC lubrication using a full atomistic model. 
In the boundary lubricated regime we show that nanoscale friction can be tuned by controlling the distribution of ion positions on muscovite mica. {\color{black} In case of commensurate setup, when ions are periodic we observe a larger friction force as compared to the case where ions are randomly placed on mica. In the latter case, the director field of the confined LCs fluctuates while  in the former case, director field orients along the grooves created between periodically arranged ions. 
When grooves  and the sliding direction are parallel, LCs exhibit higher order and the friction is lower as compared to the perpendicular case.  In case of incommensurate setup, the friction force is smaller than the commensurate setup where ions are randomly arranged on mica and the confined LCs orient along the sliding direction.} Tuning the charge distribution and modifying the surface geometry, one can reduce the friction in commensurate structures \cite{Guo}. The experimental probe of the ion arrangement \cite{bampoulis2017graphene} can open up novel directions in controlling nanoscale friction. 
{\color{black} However, at the limit of a large thickness,  surface effects disappear and we predict that effective viscosity of the confined LC and hexane exponentially decays to the bulk viscosity that exhibits a good agreement with experimental values.}

On the quest of potential lubricant from LC-hexane mixtures, our results show that increase of LC and hexane concentration in the hexane \cite{KUPCHINOV} and the liquid crystal dominated regions, respectively both lead to the reduction of friction. It suggests that instead of pure LCs addition of impurity results better lubrication and it can be thought of potential lubricant in applications.  By tuning the film composition, we also observe the possibility of controlling the stick-slip motion which is a major reason of wear in sliding surfaces.



\section*{Acknowledgement}
PKJ, MJA and LL are supported by the Academy of Finland through project no. 251748 (Centres
of Excellence Programme, 2012-2017). 
PKJ acknowledges support from the Academy of Finland FiDiPro program, project no. 13282993. 
LL acknowledges the support of the Academy of Finland
via an Academy Research Fellowship (project no. 268302).
WC is grateful to the financial support by the National Natural 
Science Foundation of China (Grant No. 11504384). We acknowledge the 
computational resources provided by the Aalto University School of Science ``Science-IT'' 
project, as well as those provided by CSC (Finland). We thank Jens Smiatek for useful discussions and suggestions.







\begin{thebibliography}{69}%
\makeatletter
\providecommand \@ifxundefined [1]{%
 \@ifx{#1\undefined}
}%
\providecommand \@ifnum [1]{%
 \ifnum #1\expandafter \@firstoftwo
 \else \expandafter \@secondoftwo
 \fi
}%
\providecommand \@ifx [1]{%
 \ifx #1\expandafter \@firstoftwo
 \else \expandafter \@secondoftwo
 \fi
}%
\providecommand \natexlab [1]{#1}%
\providecommand \enquote  [1]{``#1''}%
\providecommand \bibnamefont  [1]{#1}%
\providecommand \bibfnamefont [1]{#1}%
\providecommand \citenamefont [1]{#1}%
\providecommand \href@noop [0]{\@secondoftwo}%
\providecommand \href [0]{\begingroup \@sanitize@url \@href}%
\providecommand \@href[1]{\@@startlink{#1}\@@href}%
\providecommand \@@href[1]{\endgroup#1\@@endlink}%
\providecommand \@sanitize@url [0]{\catcode `\\12\catcode `\$12\catcode
  `\&12\catcode `\#12\catcode `\^12\catcode `\_12\catcode `\%12\relax}%
\providecommand \@@startlink[1]{}%
\providecommand \@@endlink[0]{}%
\providecommand \url  [0]{\begingroup\@sanitize@url \@url }%
\providecommand \@url [1]{\endgroup\@href {#1}{\urlprefix }}%
\providecommand \urlprefix  [0]{URL }%
\providecommand \Eprint [0]{\href }%
\providecommand \doibase [0]{http://dx.doi.org/}%
\providecommand \selectlanguage [0]{\@gobble}%
\providecommand \bibinfo  [0]{\@secondoftwo}%
\providecommand \bibfield  [0]{\@secondoftwo}%
\providecommand \translation [1]{[#1]}%
\providecommand \BibitemOpen [0]{}%
\providecommand \bibitemStop [0]{}%
\providecommand \bibitemNoStop [0]{.\EOS\space}%
\providecommand \EOS [0]{\spacefactor3000\relax}%
\providecommand \BibitemShut  [1]{\csname bibitem#1\endcsname}%
\let\auto@bib@innerbib\@empty
\bibitem [{\citenamefont {Bhushan}\ \emph {et~al.}(1995)\citenamefont
  {Bhushan}, \citenamefont {Israelachvili},\ and\ \citenamefont
  {Landman}}]{Bhushan}%
  \BibitemOpen
  \bibfield  {author} {\bibinfo {author} {\bibfnamefont {B.}~\bibnamefont
  {Bhushan}}, \bibinfo {author} {\bibfnamefont {J.~N.}\ \bibnamefont
  {Israelachvili}}, \ and\ \bibinfo {author} {\bibfnamefont {U.}~\bibnamefont
  {Landman}},\ }\href@noop {} {\bibfield  {journal} {\bibinfo  {journal}
  {Nature}\ }\textbf {\bibinfo {volume} {374}},\ \bibinfo {pages} {607}
  (\bibinfo {year} {1995})}\BibitemShut {NoStop}%
\bibitem [{\citenamefont {Park}\ \emph {et~al.}(2005)\citenamefont {Park},
  \citenamefont {Ogletree}, \citenamefont {Salmeron}, \citenamefont {Ribeiro},
  \citenamefont {Canfield}, \citenamefont {Jenks},\ and\ \citenamefont
  {Thiel}}]{Park}%
  \BibitemOpen
  \bibfield  {author} {\bibinfo {author} {\bibfnamefont {J.~Y.}\ \bibnamefont
  {Park}}, \bibinfo {author} {\bibfnamefont {D.~F.}\ \bibnamefont {Ogletree}},
  \bibinfo {author} {\bibfnamefont {M.}~\bibnamefont {Salmeron}}, \bibinfo
  {author} {\bibfnamefont {R.~A.}\ \bibnamefont {Ribeiro}}, \bibinfo {author}
  {\bibfnamefont {P.~C.}\ \bibnamefont {Canfield}}, \bibinfo {author}
  {\bibfnamefont {C.~J.}\ \bibnamefont {Jenks}}, \ and\ \bibinfo {author}
  {\bibfnamefont {P.~A.}\ \bibnamefont {Thiel}},\ }\href@noop {} {\bibfield
  {journal} {\bibinfo  {journal} {Science}\ }\textbf {\bibinfo {volume}
  {309}},\ \bibinfo {pages} {1354} (\bibinfo {year} {2005})}\BibitemShut
  {NoStop}%
\bibitem [{\citenamefont {Bhushan}(2005)}]{Bhushan1}%
  \BibitemOpen
  \bibfield  {author} {\bibinfo {author} {\bibfnamefont {B.}~\bibnamefont
  {Bhushan}},\ }\href@noop {} {\bibfield  {journal} {\bibinfo  {journal}
  {Wear}\ }\textbf {\bibinfo {volume} {259}},\ \bibinfo {pages} {1507}
  (\bibinfo {year} {2005})}\BibitemShut {NoStop}%
\bibitem [{\citenamefont {Smith}\ \emph {et~al.}(2013)\citenamefont {Smith},
  \citenamefont {Lovelock}, \citenamefont {Gosvami}, \citenamefont {Welton},\
  and\ \citenamefont {Perkin}}]{Smith}%
  \BibitemOpen
  \bibfield  {author} {\bibinfo {author} {\bibfnamefont {A.~M.}\ \bibnamefont
  {Smith}}, \bibinfo {author} {\bibfnamefont {K.~R.~J.}\ \bibnamefont
  {Lovelock}}, \bibinfo {author} {\bibfnamefont {N.~N.}\ \bibnamefont
  {Gosvami}}, \bibinfo {author} {\bibfnamefont {T.}~\bibnamefont {Welton}}, \
  and\ \bibinfo {author} {\bibfnamefont {S.}~\bibnamefont {Perkin}},\
  }\href@noop {} {\bibfield  {journal} {\bibinfo  {journal} {Phys. Chem. Chem.
  Phys.}\ }\textbf {\bibinfo {volume} {15}},\ \bibinfo {pages} {15317}
  (\bibinfo {year} {2013})}\BibitemShut {NoStop}%
\bibitem [{\citenamefont {Urbakh}\ \emph {et~al.}(2004)\citenamefont {Urbakh},
  \citenamefont {Klafter}, \citenamefont {Gourdon},\ and\ \citenamefont
  {Israelachvili}}]{Urbakh}%
  \BibitemOpen
  \bibfield  {author} {\bibinfo {author} {\bibfnamefont {M.}~\bibnamefont
  {Urbakh}}, \bibinfo {author} {\bibfnamefont {J.}~\bibnamefont {Klafter}},
  \bibinfo {author} {\bibfnamefont {D.}~\bibnamefont {Gourdon}}, \ and\
  \bibinfo {author} {\bibfnamefont {J.}~\bibnamefont {Israelachvili}},\
  }\href@noop {} {\bibfield  {journal} {\bibinfo  {journal} {Nature}\ }\textbf
  {\bibinfo {volume} {430}},\ \bibinfo {pages} {525} (\bibinfo {year}
  {2004})}\BibitemShut {NoStop}%
\bibitem [{\citenamefont {Frenken}(2006)}]{Frenken}%
  \BibitemOpen
  \bibfield  {author} {\bibinfo {author} {\bibfnamefont {J.}~\bibnamefont
  {Frenken}},\ }\href@noop {} {\bibfield  {journal} {\bibinfo  {journal} {Nat.
  Nanotechnol.}\ }\textbf {\bibinfo {volume} {1}},\ \bibinfo {pages} {20}
  (\bibinfo {year} {2006})}\BibitemShut {NoStop}%
\bibitem [{\citenamefont {Zhang}\ and\ \citenamefont
  {Chang}(2018)}]{C7NR07839K}%
  \BibitemOpen
  \bibfield  {author} {\bibinfo {author} {\bibfnamefont {H.}~\bibnamefont
  {Zhang}}\ and\ \bibinfo {author} {\bibfnamefont {T.}~\bibnamefont {Chang}},\
  }\href {\doibase 10.1039/C7NR07839K} {\bibfield  {journal} {\bibinfo
  {journal} {Nanoscale}\ }\textbf {\bibinfo {volume} {10}},\ \bibinfo {pages}
  {2447} (\bibinfo {year} {2018})}\BibitemShut {NoStop}%
\bibitem [{\citenamefont {Filippov}\ \emph {et~al.}(2010)\citenamefont
  {Filippov}, \citenamefont {Vanossi},\ and\ \citenamefont
  {Urbakh}}]{Filippov}%
  \BibitemOpen
  \bibfield  {author} {\bibinfo {author} {\bibfnamefont {A.~E.}\ \bibnamefont
  {Filippov}}, \bibinfo {author} {\bibfnamefont {A.}~\bibnamefont {Vanossi}}, \
  and\ \bibinfo {author} {\bibfnamefont {M.}~\bibnamefont {Urbakh}},\
  }\href@noop {} {\bibfield  {journal} {\bibinfo  {journal} {Phys. Rev. Lett.}\
  }\textbf {\bibinfo {volume} {104}},\ \bibinfo {pages} {074302} (\bibinfo
  {year} {2010})}\BibitemShut {NoStop}%
\bibitem [{\citenamefont {Hirano}\ \emph {et~al.}(1991)\citenamefont {Hirano},
  \citenamefont {Shinjo}, \citenamefont {Kaneko},\ and\ \citenamefont
  {Murata}}]{Hirano}%
  \BibitemOpen
  \bibfield  {author} {\bibinfo {author} {\bibfnamefont {M.}~\bibnamefont
  {Hirano}}, \bibinfo {author} {\bibfnamefont {K.}~\bibnamefont {Shinjo}},
  \bibinfo {author} {\bibfnamefont {R.}~\bibnamefont {Kaneko}}, \ and\ \bibinfo
  {author} {\bibfnamefont {Y.}~\bibnamefont {Murata}},\ }\href@noop {}
  {\bibfield  {journal} {\bibinfo  {journal} {Phys. Rev. Lett.}\ }\textbf
  {\bibinfo {volume} {67}},\ \bibinfo {pages} {2642} (\bibinfo {year}
  {1991})}\BibitemShut {NoStop}%
\bibitem [{\citenamefont {Filippov}\ \emph {et~al.}(2008)\citenamefont
  {Filippov}, \citenamefont {Dienwiebel}, \citenamefont {Frenken},
  \citenamefont {Klafter},\ and\ \citenamefont
  {Urbakh}}]{FilippovPhysRevLett.}%
  \BibitemOpen
  \bibfield  {author} {\bibinfo {author} {\bibfnamefont {A.~E.}\ \bibnamefont
  {Filippov}}, \bibinfo {author} {\bibfnamefont {M.}~\bibnamefont
  {Dienwiebel}}, \bibinfo {author} {\bibfnamefont {J.~W.~M.}\ \bibnamefont
  {Frenken}}, \bibinfo {author} {\bibfnamefont {J.}~\bibnamefont {Klafter}}, \
  and\ \bibinfo {author} {\bibfnamefont {M.}~\bibnamefont {Urbakh}},\ }\href
  {\doibase 10.1103/PhysRevLett.100.046102} {\bibfield  {journal} {\bibinfo
  {journal} {Phys. Rev. Lett.}\ }\textbf {\bibinfo {volume} {100}},\ \bibinfo
  {pages} {046102} (\bibinfo {year} {2008})}\BibitemShut {NoStop}%
\bibitem [{\citenamefont {Woods}\ \emph {et~al.}(2016)\citenamefont {Woods},
  \citenamefont {Withers}, \citenamefont {Zhu}, \citenamefont {Cao},
  \citenamefont {Yu}, \citenamefont {Kozikov}, \citenamefont {Ben~Shalom},
  \citenamefont {Morozov}, \citenamefont {van Wijk}, \citenamefont {Fasolino},
  \citenamefont {Katsnelson}, \citenamefont {Watanabe}, \citenamefont
  {Taniguchi}, \citenamefont {Geim}, \citenamefont {Mishchenko},\ and\
  \citenamefont {Novoselov}}]{WoodsNatComm}%
  \BibitemOpen
  \bibfield  {author} {\bibinfo {author} {\bibfnamefont {C.~R.}\ \bibnamefont
  {Woods}}, \bibinfo {author} {\bibfnamefont {F.}~\bibnamefont {Withers}},
  \bibinfo {author} {\bibfnamefont {M.~J.}\ \bibnamefont {Zhu}}, \bibinfo
  {author} {\bibfnamefont {Y.}~\bibnamefont {Cao}}, \bibinfo {author}
  {\bibfnamefont {G.}~\bibnamefont {Yu}}, \bibinfo {author} {\bibfnamefont
  {A.}~\bibnamefont {Kozikov}}, \bibinfo {author} {\bibfnamefont
  {M.}~\bibnamefont {Ben~Shalom}}, \bibinfo {author} {\bibfnamefont {S.~V.}\
  \bibnamefont {Morozov}}, \bibinfo {author} {\bibfnamefont {M.~M.}\
  \bibnamefont {van Wijk}}, \bibinfo {author} {\bibfnamefont {A.}~\bibnamefont
  {Fasolino}}, \bibinfo {author} {\bibfnamefont {M.~I.}\ \bibnamefont
  {Katsnelson}}, \bibinfo {author} {\bibfnamefont {K.}~\bibnamefont
  {Watanabe}}, \bibinfo {author} {\bibfnamefont {T.}~\bibnamefont {Taniguchi}},
  \bibinfo {author} {\bibfnamefont {A.~K.}\ \bibnamefont {Geim}}, \bibinfo
  {author} {\bibfnamefont {A.}~\bibnamefont {Mishchenko}}, \ and\ \bibinfo
  {author} {\bibfnamefont {K.~S.}\ \bibnamefont {Novoselov}},\ }\href@noop {}
  {\bibfield  {journal} {\bibinfo  {journal} {Nature Communications}\ }\textbf
  {\bibinfo {volume} {7}},\ \bibinfo {pages} {10800} (\bibinfo {year}
  {2016})}\BibitemShut {NoStop}%
\bibitem [{\citenamefont {Mishchenko}\ \emph {et~al.}(2014)\citenamefont
  {Mishchenko}, \citenamefont {Tu}, \citenamefont {Cao}, \citenamefont
  {Gorbachev}, \citenamefont {Wallbank}, \citenamefont {Greenaway},
  \citenamefont {Morozov}, \citenamefont {Morozov}, \citenamefont {Zhu},
  \citenamefont {Wong}, \citenamefont {Withers}, \citenamefont {Woods},
  \citenamefont {Kim}, \citenamefont {Watanabe}, \citenamefont {Taniguchi},
  \citenamefont {Vdovin}, \citenamefont {Makarovsky}, \citenamefont {Fromhold},
  \citenamefont {Fal'ko}, \citenamefont {Geim}, \citenamefont {Eaves},\ and\
  \citenamefont {Novoselov}}]{Mishchenko}%
  \BibitemOpen
  \bibfield  {author} {\bibinfo {author} {\bibfnamefont {A.}~\bibnamefont
  {Mishchenko}}, \bibinfo {author} {\bibfnamefont {J.~S.}\ \bibnamefont {Tu}},
  \bibinfo {author} {\bibfnamefont {Y.}~\bibnamefont {Cao}}, \bibinfo {author}
  {\bibfnamefont {R.~V.}\ \bibnamefont {Gorbachev}}, \bibinfo {author}
  {\bibfnamefont {J.~R.}\ \bibnamefont {Wallbank}}, \bibinfo {author}
  {\bibfnamefont {M.~T.}\ \bibnamefont {Greenaway}}, \bibinfo {author}
  {\bibfnamefont {V.~E.}\ \bibnamefont {Morozov}}, \bibinfo {author}
  {\bibfnamefont {S.~V.}\ \bibnamefont {Morozov}}, \bibinfo {author}
  {\bibfnamefont {M.~J.}\ \bibnamefont {Zhu}}, \bibinfo {author} {\bibfnamefont
  {S.~L.}\ \bibnamefont {Wong}}, \bibinfo {author} {\bibfnamefont
  {F.}~\bibnamefont {Withers}}, \bibinfo {author} {\bibfnamefont {C.~R.}\
  \bibnamefont {Woods}}, \bibinfo {author} {\bibfnamefont {Y.-J.}\ \bibnamefont
  {Kim}}, \bibinfo {author} {\bibfnamefont {K.}~\bibnamefont {Watanabe}},
  \bibinfo {author} {\bibfnamefont {T.}~\bibnamefont {Taniguchi}}, \bibinfo
  {author} {\bibfnamefont {E.~E.}\ \bibnamefont {Vdovin}}, \bibinfo {author}
  {\bibfnamefont {O.}~\bibnamefont {Makarovsky}}, \bibinfo {author}
  {\bibfnamefont {T.~M.}\ \bibnamefont {Fromhold}}, \bibinfo {author}
  {\bibfnamefont {V.~I.}\ \bibnamefont {Fal'ko}}, \bibinfo {author}
  {\bibfnamefont {A.~K.}\ \bibnamefont {Geim}}, \bibinfo {author}
  {\bibfnamefont {L.}~\bibnamefont {Eaves}}, \ and\ \bibinfo {author}
  {\bibfnamefont {K.~S.}\ \bibnamefont {Novoselov}},\ }\href
  {http://dx.doi.org/10.1038/nnano.2014.187} {\bibfield  {journal} {\bibinfo
  {journal} {Nature Nanotechnology}\ }\textbf {\bibinfo {volume} {9}},\
  \bibinfo {pages} {808} (\bibinfo {year} {2014})}\BibitemShut {NoStop}%
\bibitem [{\citenamefont {Guo}\ \emph {et~al.}(2016)\citenamefont {Guo},
  \citenamefont {Qiu},\ and\ \citenamefont {Guo}}]{Guo}%
  \BibitemOpen
  \bibfield  {author} {\bibinfo {author} {\bibfnamefont {Y.}~\bibnamefont
  {Guo}}, \bibinfo {author} {\bibfnamefont {J.}~\bibnamefont {Qiu}}, \ and\
  \bibinfo {author} {\bibfnamefont {W.}~\bibnamefont {Guo}},\ }\href@noop {}
  {\bibfield  {journal} {\bibinfo  {journal} {Nanoscale}\ }\textbf {\bibinfo
  {volume} {8}},\ \bibinfo {pages} {575} (\bibinfo {year} {2016})}\BibitemShut
  {NoStop}%
\bibitem [{\citenamefont {Thompson}\ and\ \citenamefont
  {Troian}(1997)}]{Thompson1}%
  \BibitemOpen
  \bibfield  {author} {\bibinfo {author} {\bibfnamefont {P.~A.}\ \bibnamefont
  {Thompson}}\ and\ \bibinfo {author} {\bibfnamefont {S.~M.}\ \bibnamefont
  {Troian}},\ }\href@noop {} {\bibfield  {journal} {\bibinfo  {journal}
  {Nature}\ }\textbf {\bibinfo {volume} {389}},\ \bibinfo {pages} {360}
  (\bibinfo {year} {1997})}\BibitemShut {NoStop}%
\bibitem [{\citenamefont {Zaidan}\ \emph {et~al.}(2016)\citenamefont {Zaidan},
  \citenamefont {Canova}, \citenamefont {Laurson},\ and\ \citenamefont
  {Foster}}]{zaidan2016mixture}%
  \BibitemOpen
  \bibfield  {author} {\bibinfo {author} {\bibfnamefont {M.~A.}\ \bibnamefont
  {Zaidan}}, \bibinfo {author} {\bibfnamefont {F.~F.}\ \bibnamefont {Canova}},
  \bibinfo {author} {\bibfnamefont {L.}~\bibnamefont {Laurson}}, \ and\
  \bibinfo {author} {\bibfnamefont {A.~S.}\ \bibnamefont {Foster}},\
  }\href@noop {} {\bibfield  {journal} {\bibinfo  {journal} {J. Chem. Theory
  Comput.}\ }\textbf {\bibinfo {volume} {13}},\ \bibinfo {pages} {3} (\bibinfo
  {year} {2016})}\BibitemShut {NoStop}%
\bibitem [{\citenamefont {Ouyang}\ \emph {et~al.}(2018)\citenamefont {Ouyang},
  \citenamefont {de~Wijn},\ and\ \citenamefont {Urbakh}}]{C7NR09530A}%
  \BibitemOpen
  \bibfield  {author} {\bibinfo {author} {\bibfnamefont {W.}~\bibnamefont
  {Ouyang}}, \bibinfo {author} {\bibfnamefont {A.~S.}\ \bibnamefont {de~Wijn}},
  \ and\ \bibinfo {author} {\bibfnamefont {M.}~\bibnamefont {Urbakh}},\
  }\href@noop {} {\bibfield  {journal} {\bibinfo  {journal} {Nanoscale}\
  }\textbf {\bibinfo {volume} {10}},\ \bibinfo {pages} {6375} (\bibinfo {year}
  {2018})}\BibitemShut {NoStop}%
\bibitem [{\citenamefont {Amann}\ and\ \citenamefont {Kailer}(2010)}]{Amann}%
  \BibitemOpen
  \bibfield  {author} {\bibinfo {author} {\bibfnamefont {T.}~\bibnamefont
  {Amann}}\ and\ \bibinfo {author} {\bibfnamefont {A.}~\bibnamefont {Kailer}},\
  }\href@noop {} {\bibfield  {journal} {\bibinfo  {journal} {Tribol. Lett.}\
  }\textbf {\bibinfo {volume} {37}},\ \bibinfo {pages} {343} (\bibinfo {year}
  {2010})}\BibitemShut {NoStop}%
\bibitem [{\citenamefont {Manzato}\ \emph {et~al.}(2015)\citenamefont
  {Manzato}, \citenamefont {Foster}, \citenamefont {Alava},\ and\ \citenamefont
  {Laurson}}]{Manzato}%
  \BibitemOpen
  \bibfield  {author} {\bibinfo {author} {\bibfnamefont {C.}~\bibnamefont
  {Manzato}}, \bibinfo {author} {\bibfnamefont {A.~S.}\ \bibnamefont {Foster}},
  \bibinfo {author} {\bibfnamefont {M.~J.}\ \bibnamefont {Alava}}, \ and\
  \bibinfo {author} {\bibfnamefont {L.}~\bibnamefont {Laurson}},\ }\href@noop
  {} {\bibfield  {journal} {\bibinfo  {journal} {Phys. Rev. E}\ }\textbf
  {\bibinfo {volume} {91}},\ \bibinfo {pages} {012504} (\bibinfo {year}
  {2015})}\BibitemShut {NoStop}%
\bibitem [{\citenamefont {Strelcov}\ \emph {et~al.}(2015)\citenamefont
  {Strelcov}, \citenamefont {Kumar}, \citenamefont {Bocharova}, \citenamefont
  {Sumpter}, \citenamefont {Tselev},\ and\ \citenamefont {Kalinin}}]{Strelcov}%
  \BibitemOpen
  \bibfield  {author} {\bibinfo {author} {\bibfnamefont {E.}~\bibnamefont
  {Strelcov}}, \bibinfo {author} {\bibfnamefont {R.}~\bibnamefont {Kumar}},
  \bibinfo {author} {\bibfnamefont {V.}~\bibnamefont {Bocharova}}, \bibinfo
  {author} {\bibfnamefont {B.~G.}\ \bibnamefont {Sumpter}}, \bibinfo {author}
  {\bibfnamefont {A.}~\bibnamefont {Tselev}}, \ and\ \bibinfo {author}
  {\bibfnamefont {S.~V.}\ \bibnamefont {Kalinin}},\ }\href@noop {} {\bibfield
  {journal} {\bibinfo  {journal} {Sci. Rep.}\ }\textbf {\bibinfo {volume}
  {5}},\ \bibinfo {pages} {8049 EP} (\bibinfo {year} {2015})}\BibitemShut
  {NoStop}%
\bibitem [{\citenamefont {Hu}\ \emph {et~al.}(1991)\citenamefont {Hu},
  \citenamefont {Carson},\ and\ \citenamefont {Granick}}]{Hsuan_Wei}%
  \BibitemOpen
  \bibfield  {author} {\bibinfo {author} {\bibfnamefont {H.-W.}\ \bibnamefont
  {Hu}}, \bibinfo {author} {\bibfnamefont {G.~A.}\ \bibnamefont {Carson}}, \
  and\ \bibinfo {author} {\bibfnamefont {S.}~\bibnamefont {Granick}},\
  }\href@noop {} {\bibfield  {journal} {\bibinfo  {journal} {Phys. Rev. Lett.}\
  }\textbf {\bibinfo {volume} {66}},\ \bibinfo {pages} {2758} (\bibinfo {year}
  {1991})}\BibitemShut {NoStop}%
\bibitem [{\citenamefont {Nakano}\ \emph {et~al.}(2014)\citenamefont {Nakano},
  \citenamefont {Mizukami},\ and\ \citenamefont {Kurihara}}]{Nakano}%
  \BibitemOpen
  \bibfield  {author} {\bibinfo {author} {\bibfnamefont {S.}~\bibnamefont
  {Nakano}}, \bibinfo {author} {\bibfnamefont {M.}~\bibnamefont {Mizukami}}, \
  and\ \bibinfo {author} {\bibfnamefont {K.}~\bibnamefont {Kurihara}},\
  }\href@noop {} {\bibfield  {journal} {\bibinfo  {journal} {Soft Matter}\
  }\textbf {\bibinfo {volume} {10}},\ \bibinfo {pages} {2110} (\bibinfo {year}
  {2014})}\BibitemShut {NoStop}%
\bibitem [{\citenamefont {Chen}\ \emph {et~al.}(2014)\citenamefont {Chen},
  \citenamefont {Kulju}, \citenamefont {Foster}, \citenamefont {Alava},\ and\
  \citenamefont {Laurson}}]{Chen}%
  \BibitemOpen
  \bibfield  {author} {\bibinfo {author} {\bibfnamefont {W.}~\bibnamefont
  {Chen}}, \bibinfo {author} {\bibfnamefont {S.}~\bibnamefont {Kulju}},
  \bibinfo {author} {\bibfnamefont {A.~S.}\ \bibnamefont {Foster}}, \bibinfo
  {author} {\bibfnamefont {M.~J.}\ \bibnamefont {Alava}}, \ and\ \bibinfo
  {author} {\bibfnamefont {L.}~\bibnamefont {Laurson}},\ }\href@noop {}
  {\bibfield  {journal} {\bibinfo  {journal} {Phys. Rev. E}\ }\textbf {\bibinfo
  {volume} {90}},\ \bibinfo {pages} {012404} (\bibinfo {year}
  {2014})}\BibitemShut {NoStop}%
\bibitem [{\citenamefont {MarÌa-Dolores~Berm?dez}(1997)}]{BERMUDEZ_1997}%
  \BibitemOpen
  \bibfield  {author} {\bibinfo {author} {\bibfnamefont {F.-J. C.-V.}\
  \bibnamefont {MarÌa-Dolores~Berm?dez}, \bibfnamefont {GinÈs
  MartÌnez-Nicol·s}},\ }\href@noop {} {\bibfield  {journal} {\bibinfo
  {journal} {Wear}\ }\textbf {\bibinfo {volume} {212}},\ \bibinfo {pages} {188
  } (\bibinfo {year} {1997})}\BibitemShut {NoStop}%
\bibitem [{\citenamefont {P.~Iglesias}(2004)}]{IGLESIAS_2004}%
  \BibitemOpen
  \bibfield  {author} {\bibinfo {author} {\bibfnamefont {F.~C. G. M.-N.}\
  \bibnamefont {P.~Iglesias}, \bibfnamefont {M.D.~Berm?dez}},\ }\href@noop {}
  {\bibfield  {journal} {\bibinfo  {journal} {Wear}\ }\textbf {\bibinfo
  {volume} {256}},\ \bibinfo {pages} {386 } (\bibinfo {year}
  {2004})}\BibitemShut {NoStop}%
\bibitem [{\citenamefont {Carrion}\ \emph {et~al.}(2009)\citenamefont
  {Carrion}, \citenamefont {MartÌnez-Nicolas}, \citenamefont {Iglesias},
  \citenamefont {Sanes},\ and\ \citenamefont {Bermudez}}]{Carrion_2009}%
  \BibitemOpen
  \bibfield  {author} {\bibinfo {author} {\bibfnamefont {F.-J.}\ \bibnamefont
  {Carrion}}, \bibinfo {author} {\bibfnamefont {G.}~\bibnamefont
  {MartÌnez-Nicolas}}, \bibinfo {author} {\bibfnamefont {P.}~\bibnamefont
  {Iglesias}}, \bibinfo {author} {\bibfnamefont {J.}~\bibnamefont {Sanes}}, \
  and\ \bibinfo {author} {\bibfnamefont {M.-D.}\ \bibnamefont {Bermudez}},\
  }\href@noop {} {\bibfield  {journal} {\bibinfo  {journal} {Int. J. Mol.
  Sci.}\ }\textbf {\bibinfo {volume} {10}},\ \bibinfo {pages} {4102} (\bibinfo
  {year} {2009})}\BibitemShut {NoStop}%
\bibitem [{\citenamefont {Ruths}\ \emph {et~al.}(1996)\citenamefont {Ruths},
  \citenamefont {Steinberg},\ and\ \citenamefont {Israelachvili}}]{Ruths}%
  \BibitemOpen
  \bibfield  {author} {\bibinfo {author} {\bibfnamefont {M.}~\bibnamefont
  {Ruths}}, \bibinfo {author} {\bibfnamefont {S.}~\bibnamefont {Steinberg}}, \
  and\ \bibinfo {author} {\bibfnamefont {J.~N.}\ \bibnamefont
  {Israelachvili}},\ }\href@noop {} {\bibfield  {journal} {\bibinfo  {journal}
  {Langmuir}\ }\textbf {\bibinfo {volume} {12}},\ \bibinfo {pages} {6637}
  (\bibinfo {year} {1996})}\BibitemShut {NoStop}%
\bibitem [{\citenamefont {Noirez}\ \emph {et~al.}(2005)\citenamefont {Noirez},
  \citenamefont {PÈpy},\ and\ \citenamefont {Baroni}}]{Noirez}%
  \BibitemOpen
  \bibfield  {author} {\bibinfo {author} {\bibfnamefont {L.}~\bibnamefont
  {Noirez}}, \bibinfo {author} {\bibfnamefont {G.}~\bibnamefont {PÈpy}}, \ and\
  \bibinfo {author} {\bibfnamefont {P.}~\bibnamefont {Baroni}},\ }\href@noop {}
  {\bibfield  {journal} {\bibinfo  {journal} {J. Phys. Condens. Matter}\
  }\textbf {\bibinfo {volume} {17}},\ \bibinfo {pages} {S3155} (\bibinfo {year}
  {2005})}\BibitemShut {NoStop}%
\bibitem [{\citenamefont {Bushby}\ and\ \citenamefont
  {Kawata}(2011)}]{Richard}%
  \BibitemOpen
  \bibfield  {author} {\bibinfo {author} {\bibfnamefont {R.~J.}\ \bibnamefont
  {Bushby}}\ and\ \bibinfo {author} {\bibfnamefont {K.}~\bibnamefont
  {Kawata}},\ }\href@noop {} {\bibfield  {journal} {\bibinfo  {journal} {Liq.
  Cryst.}\ }\textbf {\bibinfo {volume} {38}},\ \bibinfo {pages} {1415}
  (\bibinfo {year} {2011})}\BibitemShut {NoStop}%
\bibitem [{\citenamefont {Pujolle-Robic}\ and\ \citenamefont
  {Noirez}(2003)}]{Robic}%
  \BibitemOpen
  \bibfield  {author} {\bibinfo {author} {\bibfnamefont {C.}~\bibnamefont
  {Pujolle-Robic}}\ and\ \bibinfo {author} {\bibfnamefont {L.}~\bibnamefont
  {Noirez}},\ }\href@noop {} {\bibfield  {journal} {\bibinfo  {journal} {Phys.
  Rev. E}\ }\textbf {\bibinfo {volume} {68}},\ \bibinfo {pages} {061706}
  (\bibinfo {year} {2003})}\BibitemShut {NoStop}%
\bibitem [{\citenamefont {Idziak}\ \emph {et~al.}(1994)\citenamefont {Idziak},
  \citenamefont {Safinya}, \citenamefont {Hill}, \citenamefont {Kraiser},
  \citenamefont {Ruths}, \citenamefont {Warriner}, \citenamefont {Steinberg},
  \citenamefont {Liang},\ and\ \citenamefont {Israelachvili}}]{Idziak1915}%
  \BibitemOpen
  \bibfield  {author} {\bibinfo {author} {\bibfnamefont {S.~H.}\ \bibnamefont
  {Idziak}}, \bibinfo {author} {\bibfnamefont {C.~R.}\ \bibnamefont {Safinya}},
  \bibinfo {author} {\bibfnamefont {R.~S.}\ \bibnamefont {Hill}}, \bibinfo
  {author} {\bibfnamefont {K.~E.}\ \bibnamefont {Kraiser}}, \bibinfo {author}
  {\bibfnamefont {M.}~\bibnamefont {Ruths}}, \bibinfo {author} {\bibfnamefont
  {H.~E.}\ \bibnamefont {Warriner}}, \bibinfo {author} {\bibfnamefont
  {S.}~\bibnamefont {Steinberg}}, \bibinfo {author} {\bibfnamefont {K.~S.}\
  \bibnamefont {Liang}}, \ and\ \bibinfo {author} {\bibfnamefont {J.~N.}\
  \bibnamefont {Israelachvili}},\ }\href@noop {} {\bibfield  {journal}
  {\bibinfo  {journal} {Science}\ }\textbf {\bibinfo {volume} {264}},\ \bibinfo
  {pages} {1915} (\bibinfo {year} {1994})}\BibitemShut {NoStop}%
\bibitem [{\citenamefont {Cheng}\ \emph {et~al.}(2008)\citenamefont {Cheng},
  \citenamefont {Kellogg}, \citenamefont {Shkoller},\ and\ \citenamefont
  {Turcotte}}]{Cheng10062008}%
  \BibitemOpen
  \bibfield  {author} {\bibinfo {author} {\bibfnamefont {C.~H.~A.}\
  \bibnamefont {Cheng}}, \bibinfo {author} {\bibfnamefont {L.~H.}\ \bibnamefont
  {Kellogg}}, \bibinfo {author} {\bibfnamefont {S.}~\bibnamefont {Shkoller}}, \
  and\ \bibinfo {author} {\bibfnamefont {D.~L.}\ \bibnamefont {Turcotte}},\
  }\href {\doibase 10.1073/pnas.0710990105} {\bibfield  {journal} {\bibinfo
  {journal} {Proceedings of the National Academy of Sciences}\ }\textbf
  {\bibinfo {volume} {105}},\ \bibinfo {pages} {7930} (\bibinfo {year}
  {2008})}\BibitemShut {NoStop}%
\bibitem [{\citenamefont {Zhang}\ \emph {et~al.}(2015)\citenamefont {Zhang},
  \citenamefont {Zhang}, \citenamefont {Qiao}, \citenamefont {Guo},
  \citenamefont {Tian},\ and\ \citenamefont {Meng}}]{Zhang_2015}%
  \BibitemOpen
  \bibfield  {author} {\bibinfo {author} {\bibfnamefont {X.}~\bibnamefont
  {Zhang}}, \bibinfo {author} {\bibfnamefont {X.}~\bibnamefont {Zhang}},
  \bibinfo {author} {\bibfnamefont {X.}~\bibnamefont {Qiao}}, \bibinfo {author}
  {\bibfnamefont {Y.}~\bibnamefont {Guo}}, \bibinfo {author} {\bibfnamefont
  {Y.}~\bibnamefont {Tian}}, \ and\ \bibinfo {author} {\bibfnamefont
  {Y.}~\bibnamefont {Meng}},\ }\href@noop {} {\bibfield  {journal} {\bibinfo
  {journal} {Microfluidics and Nanofluidics}\ }\textbf {\bibinfo {volume}
  {18}},\ \bibinfo {pages} {1131} (\bibinfo {year} {2015})}\BibitemShut
  {NoStop}%
\bibitem [{\citenamefont {Cifelli}\ \emph {et~al.}(2008)\citenamefont
  {Cifelli}, \citenamefont {De~Gaetani}, \citenamefont {Prampolini},\ and\
  \citenamefont {Tani}}]{cifelli2008atomistic}%
  \BibitemOpen
  \bibfield  {author} {\bibinfo {author} {\bibfnamefont {M.}~\bibnamefont
  {Cifelli}}, \bibinfo {author} {\bibfnamefont {L.}~\bibnamefont {De~Gaetani}},
  \bibinfo {author} {\bibfnamefont {G.}~\bibnamefont {Prampolini}}, \ and\
  \bibinfo {author} {\bibfnamefont {A.}~\bibnamefont {Tani}},\ }\href@noop {}
  {\bibfield  {journal} {\bibinfo  {journal} {The Journal of Physical Chemistry
  B}\ }\textbf {\bibinfo {volume} {112}},\ \bibinfo {pages} {9777} (\bibinfo
  {year} {2008})}\BibitemShut {NoStop}%
\bibitem [{\citenamefont {Tiberio}\ \emph {et~al.}(2009)\citenamefont
  {Tiberio}, \citenamefont {Muccioli}, \citenamefont {Berardi},\ and\
  \citenamefont {Zannoni}}]{tiberio2009towards}%
  \BibitemOpen
  \bibfield  {author} {\bibinfo {author} {\bibfnamefont {G.}~\bibnamefont
  {Tiberio}}, \bibinfo {author} {\bibfnamefont {L.}~\bibnamefont {Muccioli}},
  \bibinfo {author} {\bibfnamefont {R.}~\bibnamefont {Berardi}}, \ and\
  \bibinfo {author} {\bibfnamefont {C.}~\bibnamefont {Zannoni}},\ }\href@noop
  {} {\bibfield  {journal} {\bibinfo  {journal} {ChemPhysChem}\ }\textbf
  {\bibinfo {volume} {10}},\ \bibinfo {pages} {125} (\bibinfo {year}
  {2009})}\BibitemShut {NoStop}%
\bibitem [{\citenamefont {Janik}\ \emph {et~al.}(1997)\citenamefont {Janik},
  \citenamefont {Tadmor},\ and\ \citenamefont {Klein}}]{janik1997shear}%
  \BibitemOpen
  \bibfield  {author} {\bibinfo {author} {\bibfnamefont {J.}~\bibnamefont
  {Janik}}, \bibinfo {author} {\bibfnamefont {R.}~\bibnamefont {Tadmor}}, \
  and\ \bibinfo {author} {\bibfnamefont {J.}~\bibnamefont {Klein}},\
  }\href@noop {} {\bibfield  {journal} {\bibinfo  {journal} {Langmuir}\
  }\textbf {\bibinfo {volume} {13}},\ \bibinfo {pages} {4466} (\bibinfo {year}
  {1997})}\BibitemShut {NoStop}%
\bibitem [{\citenamefont {Kitaev}\ and\ \citenamefont
  {Kumacheva}(2000)}]{kitaev2000thin}%
  \BibitemOpen
  \bibfield  {author} {\bibinfo {author} {\bibfnamefont {V.}~\bibnamefont
  {Kitaev}}\ and\ \bibinfo {author} {\bibfnamefont {E.}~\bibnamefont
  {Kumacheva}},\ }\href@noop {} {\bibfield  {journal} {\bibinfo  {journal} {The
  Journal of Physical Chemistry B}\ }\textbf {\bibinfo {volume} {104}},\
  \bibinfo {pages} {8822} (\bibinfo {year} {2000})}\BibitemShut {NoStop}%
\bibitem [{\citenamefont {Mizukami}\ \emph {et~al.}(2004)\citenamefont
  {Mizukami}, \citenamefont {Kusakabe},\ and\ \citenamefont
  {Kurihara}}]{mizukami2004shear}%
  \BibitemOpen
  \bibfield  {author} {\bibinfo {author} {\bibfnamefont {M.}~\bibnamefont
  {Mizukami}}, \bibinfo {author} {\bibfnamefont {K.}~\bibnamefont {Kusakabe}},
  \ and\ \bibinfo {author} {\bibfnamefont {K.}~\bibnamefont {Kurihara}},\ }in\
  \href@noop {} {\emph {\bibinfo {booktitle} {Surface and Colloid Science}}}\
  (\bibinfo  {publisher} {Springer},\ \bibinfo {year} {2004})\ pp.\ \bibinfo
  {pages} {105--108}\BibitemShut {NoStop}%
\bibitem [{\citenamefont {Adam}\ \emph {et~al.}(1997)\citenamefont {Adam},
  \citenamefont {Clark}, \citenamefont {Ackland},\ and\ \citenamefont
  {Crain}}]{Adam}%
  \BibitemOpen
  \bibfield  {author} {\bibinfo {author} {\bibfnamefont {C.~J.}\ \bibnamefont
  {Adam}}, \bibinfo {author} {\bibfnamefont {S.~J.}\ \bibnamefont {Clark}},
  \bibinfo {author} {\bibfnamefont {G.~J.}\ \bibnamefont {Ackland}}, \ and\
  \bibinfo {author} {\bibfnamefont {J.}~\bibnamefont {Crain}},\ }\href@noop {}
  {\bibfield  {journal} {\bibinfo  {journal} {Phys. Rev. E}\ }\textbf {\bibinfo
  {volume} {55}},\ \bibinfo {pages} {5641} (\bibinfo {year}
  {1997})}\BibitemShut {NoStop}%
\bibitem [{\citenamefont {Cheung}\ \emph {et~al.}(2002)\citenamefont {Cheung},
  \citenamefont {Clark},\ and\ \citenamefont {Wilson}}]{Cheung}%
  \BibitemOpen
  \bibfield  {author} {\bibinfo {author} {\bibfnamefont {D.~L.}\ \bibnamefont
  {Cheung}}, \bibinfo {author} {\bibfnamefont {S.~J.}\ \bibnamefont {Clark}}, \
  and\ \bibinfo {author} {\bibfnamefont {M.~R.}\ \bibnamefont {Wilson}},\
  }\href@noop {} {\bibfield  {journal} {\bibinfo  {journal} {Phys. Rev. E}\
  }\textbf {\bibinfo {volume} {65}},\ \bibinfo {pages} {051709} (\bibinfo
  {year} {2002})}\BibitemShut {NoStop}%
\bibitem [{\citenamefont {Heinz}\ \emph {et~al.}(2005)\citenamefont {Heinz},
  \citenamefont {Koerner}, \citenamefont {Anderson}, \citenamefont {Vaia},\
  and\ \citenamefont {Farmer}}]{Heinz}%
  \BibitemOpen
  \bibfield  {author} {\bibinfo {author} {\bibfnamefont {H.}~\bibnamefont
  {Heinz}}, \bibinfo {author} {\bibfnamefont {H.}~\bibnamefont {Koerner}},
  \bibinfo {author} {\bibfnamefont {K.~L.}\ \bibnamefont {Anderson}}, \bibinfo
  {author} {\bibfnamefont {R.~A.}\ \bibnamefont {Vaia}}, \ and\ \bibinfo
  {author} {\bibfnamefont {B.~L.}\ \bibnamefont {Farmer}},\ }\href@noop {}
  {\bibfield  {journal} {\bibinfo  {journal} {Chem. Mater.}\ }\textbf {\bibinfo
  {volume} {17}},\ \bibinfo {pages} {5658} (\bibinfo {year}
  {2005})}\BibitemShut {NoStop}%
\bibitem [{\citenamefont {Xu}\ and\ \citenamefont
  {Salmeron}(1998)}]{xu1998effects}%
  \BibitemOpen
  \bibfield  {author} {\bibinfo {author} {\bibfnamefont {L.}~\bibnamefont
  {Xu}}\ and\ \bibinfo {author} {\bibfnamefont {M.}~\bibnamefont {Salmeron}},\
  }\href@noop {} {\bibfield  {journal} {\bibinfo  {journal} {Langmuir}\
  }\textbf {\bibinfo {volume} {14}},\ \bibinfo {pages} {2187} (\bibinfo {year}
  {1998})}\BibitemShut {NoStop}%
\bibitem [{\citenamefont {Ricci}\ \emph {et~al.}(2014)\citenamefont {Ricci},
  \citenamefont {Spijker},\ and\ \citenamefont
  {Vo{\"\i}tchovsky}}]{ricci2014water}%
  \BibitemOpen
  \bibfield  {author} {\bibinfo {author} {\bibfnamefont {M.}~\bibnamefont
  {Ricci}}, \bibinfo {author} {\bibfnamefont {P.}~\bibnamefont {Spijker}}, \
  and\ \bibinfo {author} {\bibfnamefont {K.}~\bibnamefont {Vo{\"\i}tchovsky}},\
  }\href@noop {} {\bibfield  {journal} {\bibinfo  {journal} {Nature
  communications}\ }\textbf {\bibinfo {volume} {5}},\ \bibinfo {pages} {4400}
  (\bibinfo {year} {2014})}\BibitemShut {NoStop}%
\bibitem [{\citenamefont {Odelius}\ \emph {et~al.}(1997)\citenamefont
  {Odelius}, \citenamefont {Bernasconi},\ and\ \citenamefont
  {Parrinello}}]{Odelius}%
  \BibitemOpen
  \bibfield  {author} {\bibinfo {author} {\bibfnamefont {M.}~\bibnamefont
  {Odelius}}, \bibinfo {author} {\bibfnamefont {M.}~\bibnamefont {Bernasconi}},
  \ and\ \bibinfo {author} {\bibfnamefont {M.}~\bibnamefont {Parrinello}},\
  }\href@noop {} {\bibfield  {journal} {\bibinfo  {journal} {Phys. Rev. Lett.}\
  }\textbf {\bibinfo {volume} {78}},\ \bibinfo {pages} {2855} (\bibinfo {year}
  {1997})}\BibitemShut {NoStop}%
\bibitem [{\citenamefont {Bampoulis}\ \emph {et~al.}(2017)\citenamefont
  {Bampoulis}, \citenamefont {Sotthewes}, \citenamefont {Siekman},
  \citenamefont {Zandvliet},\ and\ \citenamefont
  {Poelsema}}]{bampoulis2017graphene}%
  \BibitemOpen
  \bibfield  {author} {\bibinfo {author} {\bibfnamefont {P.}~\bibnamefont
  {Bampoulis}}, \bibinfo {author} {\bibfnamefont {K.}~\bibnamefont
  {Sotthewes}}, \bibinfo {author} {\bibfnamefont {M.~H.}\ \bibnamefont
  {Siekman}}, \bibinfo {author} {\bibfnamefont {H.~J.}\ \bibnamefont
  {Zandvliet}}, \ and\ \bibinfo {author} {\bibfnamefont {B.}~\bibnamefont
  {Poelsema}},\ }\href@noop {} {\bibfield  {journal} {\bibinfo  {journal} {Sci.
  Rep.}\ }\textbf {\bibinfo {volume} {7}},\ \bibinfo {pages} {43451} (\bibinfo
  {year} {2017})}\BibitemShut {NoStop}%
\bibitem [{\citenamefont {Ouyang}\ \emph {et~al.}(2016)\citenamefont {Ouyang},
  \citenamefont {Ma}, \citenamefont {Zheng},\ and\ \citenamefont
  {Urbakh}}]{Ouyang}%
  \BibitemOpen
  \bibfield  {author} {\bibinfo {author} {\bibfnamefont {W.}~\bibnamefont
  {Ouyang}}, \bibinfo {author} {\bibfnamefont {M.}~\bibnamefont {Ma}}, \bibinfo
  {author} {\bibfnamefont {Q.}~\bibnamefont {Zheng}}, \ and\ \bibinfo {author}
  {\bibfnamefont {M.}~\bibnamefont {Urbakh}},\ }\href@noop {} {\bibfield
  {journal} {\bibinfo  {journal} {Nano Letters}\ }\textbf {\bibinfo {volume}
  {16}},\ \bibinfo {pages} {1878} (\bibinfo {year} {2016})}\BibitemShut
  {NoStop}%
\bibitem [{\citenamefont {Dienwiebel}\ \emph {et~al.}(2004)\citenamefont
  {Dienwiebel}, \citenamefont {Verhoeven}, \citenamefont {Pradeep},
  \citenamefont {Frenken}, \citenamefont {Heimberg},\ and\ \citenamefont
  {Zandbergen}}]{Dienwiebel}%
  \BibitemOpen
  \bibfield  {author} {\bibinfo {author} {\bibfnamefont {M.}~\bibnamefont
  {Dienwiebel}}, \bibinfo {author} {\bibfnamefont {G.~S.}\ \bibnamefont
  {Verhoeven}}, \bibinfo {author} {\bibfnamefont {N.}~\bibnamefont {Pradeep}},
  \bibinfo {author} {\bibfnamefont {J.~W.~M.}\ \bibnamefont {Frenken}},
  \bibinfo {author} {\bibfnamefont {J.~A.}\ \bibnamefont {Heimberg}}, \ and\
  \bibinfo {author} {\bibfnamefont {H.~W.}\ \bibnamefont {Zandbergen}},\
  }\href@noop {} {\bibfield  {journal} {\bibinfo  {journal} {Phys. Rev. Lett.}\
  }\textbf {\bibinfo {volume} {92}},\ \bibinfo {pages} {126101} (\bibinfo
  {year} {2004})}\BibitemShut {NoStop}%
\bibitem [{\citenamefont {Yoshizawa}\ and\ \citenamefont
  {Israelachvili}(1993)}]{Yoshizawa}%
  \BibitemOpen
  \bibfield  {author} {\bibinfo {author} {\bibfnamefont {H.}~\bibnamefont
  {Yoshizawa}}\ and\ \bibinfo {author} {\bibfnamefont {J.}~\bibnamefont
  {Israelachvili}},\ }\href@noop {} {\bibfield  {journal} {\bibinfo  {journal}
  {The Journal of Physical Chemistry}\ }\textbf {\bibinfo {volume} {97}},\
  \bibinfo {pages} {11300} (\bibinfo {year} {1993})}\BibitemShut {NoStop}%
\bibitem [{\citenamefont {Lei}\ and\ \citenamefont {Leng}(2011)}]{Lei}%
  \BibitemOpen
  \bibfield  {author} {\bibinfo {author} {\bibfnamefont {Y.}~\bibnamefont
  {Lei}}\ and\ \bibinfo {author} {\bibfnamefont {Y.}~\bibnamefont {Leng}},\
  }\href@noop {} {\bibfield  {journal} {\bibinfo  {journal} {Phys. Rev. Lett.}\
  }\textbf {\bibinfo {volume} {107}},\ \bibinfo {pages} {147801} (\bibinfo
  {year} {2011})}\BibitemShut {NoStop}%
\bibitem [{\citenamefont {Thompson}\ and\ \citenamefont
  {Robbins}(1990)}]{thompson1990shear}%
  \BibitemOpen
  \bibfield  {author} {\bibinfo {author} {\bibfnamefont {P.~A.}\ \bibnamefont
  {Thompson}}\ and\ \bibinfo {author} {\bibfnamefont {M.~O.}\ \bibnamefont
  {Robbins}},\ }\href@noop {} {\bibfield  {journal} {\bibinfo  {journal} {Phys.
  Rev. A}\ }\textbf {\bibinfo {volume} {41}},\ \bibinfo {pages} {6830}
  (\bibinfo {year} {1990})}\BibitemShut {NoStop}%
\bibitem [{\citenamefont {Chen}\ \emph {et~al.}(2015)\citenamefont {Chen},
  \citenamefont {Foster}, \citenamefont {Alava},\ and\ \citenamefont
  {Laurson}}]{chen2015stick}%
  \BibitemOpen
  \bibfield  {author} {\bibinfo {author} {\bibfnamefont {W.}~\bibnamefont
  {Chen}}, \bibinfo {author} {\bibfnamefont {A.~S.}\ \bibnamefont {Foster}},
  \bibinfo {author} {\bibfnamefont {M.~J.}\ \bibnamefont {Alava}}, \ and\
  \bibinfo {author} {\bibfnamefont {L.}~\bibnamefont {Laurson}},\ }\href@noop
  {} {\bibfield  {journal} {\bibinfo  {journal} {Phys. Rev. Lett.}\ }\textbf
  {\bibinfo {volume} {114}},\ \bibinfo {pages} {095502} (\bibinfo {year}
  {2015})}\BibitemShut {NoStop}%
\bibitem [{\citenamefont {Plimpton}(1995)}]{plimpton1995fast}%
  \BibitemOpen
  \bibfield  {author} {\bibinfo {author} {\bibfnamefont {S.}~\bibnamefont
  {Plimpton}},\ }\href@noop {} {\bibfield  {journal} {\bibinfo  {journal}
  {Journal of computational physics}\ }\textbf {\bibinfo {volume} {117}},\
  \bibinfo {pages} {1} (\bibinfo {year} {1995})}\BibitemShut {NoStop}%
\bibitem [{\citenamefont {Patel}\ \emph {et~al.}(2006)\citenamefont {Patel},
  \citenamefont {Jeon}, \citenamefont {Mather},\ and\ \citenamefont
  {Dobrynin}}]{patel2006molecular}%
  \BibitemOpen
  \bibfield  {author} {\bibinfo {author} {\bibfnamefont {P.~A.}\ \bibnamefont
  {Patel}}, \bibinfo {author} {\bibfnamefont {J.}~\bibnamefont {Jeon}},
  \bibinfo {author} {\bibfnamefont {P.~T.}\ \bibnamefont {Mather}}, \ and\
  \bibinfo {author} {\bibfnamefont {A.~V.}\ \bibnamefont {Dobrynin}},\
  }\href@noop {} {\bibfield  {journal} {\bibinfo  {journal} {Langmuir}\
  }\textbf {\bibinfo {volume} {22}},\ \bibinfo {pages} {9994} (\bibinfo {year}
  {2006})}\BibitemShut {NoStop}%
\bibitem [{\citenamefont {Pan}\ \emph {et~al.}(2017)\citenamefont {Pan},
  \citenamefont {Yi},\ and\ \citenamefont {Hu}}]{pan2017effect}%
  \BibitemOpen
  \bibfield  {author} {\bibinfo {author} {\bibfnamefont {C.}~\bibnamefont
  {Pan}}, \bibinfo {author} {\bibfnamefont {S.}~\bibnamefont {Yi}}, \ and\
  \bibinfo {author} {\bibfnamefont {Z.}~\bibnamefont {Hu}},\ }\href@noop {}
  {\bibfield  {journal} {\bibinfo  {journal} {Physical Chemistry Chemical
  Physics}\ }\textbf {\bibinfo {volume} {19}},\ \bibinfo {pages} {4861}
  (\bibinfo {year} {2017})}\BibitemShut {NoStop}%
\bibitem [{\citenamefont {Cerda}\ \emph {et~al.}(2009)\citenamefont {Cerda},
  \citenamefont {Qiao},\ and\ \citenamefont {Holm}}]{cerda2009understanding}%
  \BibitemOpen
  \bibfield  {author} {\bibinfo {author} {\bibfnamefont {J.~J.}\ \bibnamefont
  {Cerda}}, \bibinfo {author} {\bibfnamefont {B.}~\bibnamefont {Qiao}}, \ and\
  \bibinfo {author} {\bibfnamefont {C.}~\bibnamefont {Holm}},\ }\href@noop {}
  {\bibfield  {journal} {\bibinfo  {journal} {Soft Matter}\ }\textbf {\bibinfo
  {volume} {5}},\ \bibinfo {pages} {4412} (\bibinfo {year} {2009})}\BibitemShut
  {NoStop}%
\bibitem [{\citenamefont {Kumacheva}\ and\ \citenamefont
  {Klein}(1998)}]{kumacheva1998simple}%
  \BibitemOpen
  \bibfield  {author} {\bibinfo {author} {\bibfnamefont {E.}~\bibnamefont
  {Kumacheva}}\ and\ \bibinfo {author} {\bibfnamefont {J.}~\bibnamefont
  {Klein}},\ }\href@noop {} {\bibfield  {journal} {\bibinfo  {journal} {The
  Journal of chemical physics}\ }\textbf {\bibinfo {volume} {108}},\ \bibinfo
  {pages} {7010} (\bibinfo {year} {1998})}\BibitemShut {NoStop}%
\bibitem [{\citenamefont {Ohnishi}\ \emph {et~al.}(2007)\citenamefont
  {Ohnishi}, \citenamefont {Kaneko}, \citenamefont {Gong}, \citenamefont
  {Osada}, \citenamefont {Stewart},\ and\ \citenamefont
  {Yaminsky}}]{ohnishi2007influence}%
  \BibitemOpen
  \bibfield  {author} {\bibinfo {author} {\bibfnamefont {S.}~\bibnamefont
  {Ohnishi}}, \bibinfo {author} {\bibfnamefont {D.}~\bibnamefont {Kaneko}},
  \bibinfo {author} {\bibfnamefont {J.~P.}\ \bibnamefont {Gong}}, \bibinfo
  {author} {\bibfnamefont {Y.}~\bibnamefont {Osada}}, \bibinfo {author}
  {\bibfnamefont {A.~M.}\ \bibnamefont {Stewart}}, \ and\ \bibinfo {author}
  {\bibfnamefont {V.~V.}\ \bibnamefont {Yaminsky}},\ }\href@noop {} {\bibfield
  {journal} {\bibinfo  {journal} {Langmuir}\ }\textbf {\bibinfo {volume}
  {23}},\ \bibinfo {pages} {7032} (\bibinfo {year} {2007})}\BibitemShut
  {NoStop}%
\bibitem [{\citenamefont {Rosenhek-Goldian}\ \emph {et~al.}(2015)\citenamefont
  {Rosenhek-Goldian}, \citenamefont {Kampf}, \citenamefont {Yeredor},\ and\
  \citenamefont {Klein}}]{rosenhek2015question}%
  \BibitemOpen
  \bibfield  {author} {\bibinfo {author} {\bibfnamefont {I.}~\bibnamefont
  {Rosenhek-Goldian}}, \bibinfo {author} {\bibfnamefont {N.}~\bibnamefont
  {Kampf}}, \bibinfo {author} {\bibfnamefont {A.}~\bibnamefont {Yeredor}}, \
  and\ \bibinfo {author} {\bibfnamefont {J.}~\bibnamefont {Klein}},\
  }\href@noop {} {\bibfield  {journal} {\bibinfo  {journal} {Proceedings of the
  National Academy of Sciences}\ }\textbf {\bibinfo {volume} {112}},\ \bibinfo
  {pages} {7117} (\bibinfo {year} {2015})}\BibitemShut {NoStop}%
\bibitem [{\citenamefont {Van~der Vegt}\ \emph {et~al.}(2001)\citenamefont
  {Van~der Vegt}, \citenamefont {M{\"u}ller-Plathe}, \citenamefont
  {Gele{\ss}us},\ and\ \citenamefont {Johannsmann}}]{van2001orientation}%
  \BibitemOpen
  \bibfield  {author} {\bibinfo {author} {\bibfnamefont {N.}~\bibnamefont
  {Van~der Vegt}}, \bibinfo {author} {\bibfnamefont {F.}~\bibnamefont
  {M{\"u}ller-Plathe}}, \bibinfo {author} {\bibfnamefont {A.}~\bibnamefont
  {Gele{\ss}us}}, \ and\ \bibinfo {author} {\bibfnamefont {D.}~\bibnamefont
  {Johannsmann}},\ }\href@noop {} {\bibfield  {journal} {\bibinfo  {journal}
  {The Journal of Chemical Physics}\ }\textbf {\bibinfo {volume} {115}},\
  \bibinfo {pages} {9935} (\bibinfo {year} {2001})}\BibitemShut {NoStop}%
\bibitem [{\citenamefont {Hirano}\ and\ \citenamefont {Sakai}(2008)}]{Hirano2}%
  \BibitemOpen
  \bibfield  {author} {\bibinfo {author} {\bibfnamefont {T.}~\bibnamefont
  {Hirano}}\ and\ \bibinfo {author} {\bibfnamefont {K.}~\bibnamefont {Sakai}},\
  }\href@noop {} {\bibfield  {journal} {\bibinfo  {journal} {Phys. Rev. E}\
  }\textbf {\bibinfo {volume} {77}},\ \bibinfo {pages} {011703} (\bibinfo
  {year} {2008})}\BibitemShut {NoStop}%
\bibitem [{\citenamefont {Jad{\.z}yn}\ \emph {et~al.}(2000)\citenamefont
  {Jad{\.z}yn}, \citenamefont {Czechowski},\ and\ \citenamefont
  {Bauman}}]{Jan}%
  \BibitemOpen
  \bibfield  {author} {\bibinfo {author} {\bibfnamefont {J.}~\bibnamefont
  {Jad{\.z}yn}}, \bibinfo {author} {\bibfnamefont {G.}~\bibnamefont
  {Czechowski}}, \ and\ \bibinfo {author} {\bibfnamefont {D.}~\bibnamefont
  {Bauman}},\ }\href@noop {} {\bibfield  {journal} {\bibinfo  {journal} {Z.
  Naturforsch. A}\ }\textbf {\bibinfo {volume} {55}},\ \bibinfo {pages} {810}
  (\bibinfo {year} {2000})}\BibitemShut {NoStop}%
\bibitem [{\citenamefont {Rodr{\'\i}guez}\ \emph {et~al.}(2006)\citenamefont
  {Rodr{\'\i}guez}, \citenamefont {Pereiro}, \citenamefont {Canosa},\ and\
  \citenamefont {Tojo}}]{rodriguez2006dynamic}%
  \BibitemOpen
  \bibfield  {author} {\bibinfo {author} {\bibfnamefont {A.}~\bibnamefont
  {Rodr{\'\i}guez}}, \bibinfo {author} {\bibfnamefont {A.}~\bibnamefont
  {Pereiro}}, \bibinfo {author} {\bibfnamefont {J.}~\bibnamefont {Canosa}}, \
  and\ \bibinfo {author} {\bibfnamefont {J.}~\bibnamefont {Tojo}},\ }\href@noop
  {} {\bibfield  {journal} {\bibinfo  {journal} {J. Chem. Thermodyn.}\ }\textbf
  {\bibinfo {volume} {38}},\ \bibinfo {pages} {505} (\bibinfo {year}
  {2006})}\BibitemShut {NoStop}%
\bibitem [{\citenamefont {Leng}\ and\ \citenamefont {Cummings}(2005)}]{Leng}%
  \BibitemOpen
  \bibfield  {author} {\bibinfo {author} {\bibfnamefont {Y.}~\bibnamefont
  {Leng}}\ and\ \bibinfo {author} {\bibfnamefont {P.~T.}\ \bibnamefont
  {Cummings}},\ }\href@noop {} {\bibfield  {journal} {\bibinfo  {journal}
  {Phys. Rev. Lett.}\ }\textbf {\bibinfo {volume} {94}},\ \bibinfo {pages}
  {026101} (\bibinfo {year} {2005})}\BibitemShut {NoStop}%
\bibitem [{\citenamefont {Jabbarzadeh}\ \emph {et~al.}(2005)\citenamefont
  {Jabbarzadeh}, \citenamefont {Harrowell},\ and\ \citenamefont
  {Tanner}}]{jabbarzadeh2005very}%
  \BibitemOpen
  \bibfield  {author} {\bibinfo {author} {\bibfnamefont {A.}~\bibnamefont
  {Jabbarzadeh}}, \bibinfo {author} {\bibfnamefont {P.}~\bibnamefont
  {Harrowell}}, \ and\ \bibinfo {author} {\bibfnamefont {R.}~\bibnamefont
  {Tanner}},\ }\href@noop {} {\bibfield  {journal} {\bibinfo  {journal} {Phys.
  Rev. Lett.}\ }\textbf {\bibinfo {volume} {94}},\ \bibinfo {pages} {126103}
  (\bibinfo {year} {2005})}\BibitemShut {NoStop}%
\bibitem [{\citenamefont {Ratna}\ and\ \citenamefont
  {Shashidhar}(1976)}]{Ratna}%
  \BibitemOpen
  \bibfield  {author} {\bibinfo {author} {\bibfnamefont {B.~R.}\ \bibnamefont
  {Ratna}}\ and\ \bibinfo {author} {\bibfnamefont {R.}~\bibnamefont
  {Shashidhar}},\ }\href@noop {} {\bibfield  {journal} {\bibinfo  {journal}
  {Pramana}\ }\textbf {\bibinfo {volume} {6}},\ \bibinfo {pages} {278}
  (\bibinfo {year} {1976})}\BibitemShut {NoStop}%
\bibitem [{\citenamefont {Mopsik}(1967)}]{Frederick}%
  \BibitemOpen
  \bibfield  {author} {\bibinfo {author} {\bibfnamefont {F.~I.}\ \bibnamefont
  {Mopsik}},\ }\href@noop {} {\bibfield  {journal} {\bibinfo  {journal} {J.
  Res. Natl. Bur. Stand. Sec. A}\ }\textbf {\bibinfo {volume} {71 A}},\
  \bibinfo {pages} {287} (\bibinfo {year} {1967})}\BibitemShut {NoStop}%
\bibitem [{\citenamefont {Lee}\ \emph {et~al.}(2017)\citenamefont {Lee},
  \citenamefont {Perez-Martinez}, \citenamefont {Smith},\ and\ \citenamefont
  {Perkin}}]{lee2017scaling}%
  \BibitemOpen
  \bibfield  {author} {\bibinfo {author} {\bibfnamefont {A.~A.}\ \bibnamefont
  {Lee}}, \bibinfo {author} {\bibfnamefont {C.~S.}\ \bibnamefont
  {Perez-Martinez}}, \bibinfo {author} {\bibfnamefont {A.~M.}\ \bibnamefont
  {Smith}}, \ and\ \bibinfo {author} {\bibfnamefont {S.}~\bibnamefont
  {Perkin}},\ }\href {\doibase 10.1103/PhysRevLett.119.026002} {\bibfield
  {journal} {\bibinfo  {journal} {Phys. Rev. Lett.}\ }\textbf {\bibinfo
  {volume} {119}},\ \bibinfo {pages} {026002} (\bibinfo {year}
  {2017})}\BibitemShut {NoStop}%
\bibitem [{\citenamefont {Shen}\ \emph {et~al.}(2001)\citenamefont {Shen},
  \citenamefont {Luo}, \citenamefont {Wen},\ and\ \citenamefont
  {Yao}}]{shen2001nano}%
  \BibitemOpen
  \bibfield  {author} {\bibinfo {author} {\bibfnamefont {M.}~\bibnamefont
  {Shen}}, \bibinfo {author} {\bibfnamefont {J.}~\bibnamefont {Luo}}, \bibinfo
  {author} {\bibfnamefont {S.}~\bibnamefont {Wen}}, \ and\ \bibinfo {author}
  {\bibfnamefont {J.}~\bibnamefont {Yao}},\ }\href@noop {} {\bibfield
  {journal} {\bibinfo  {journal} {Chin. Sci. Bull.}\ }\textbf {\bibinfo
  {volume} {46}},\ \bibinfo {pages} {1227} (\bibinfo {year}
  {2001})}\BibitemShut {NoStop}%
\bibitem [{\citenamefont {Ruths}\ and\ \citenamefont
  {Granick}(2000)}]{ruths2000influence}%
  \BibitemOpen
  \bibfield  {author} {\bibinfo {author} {\bibfnamefont {M.}~\bibnamefont
  {Ruths}}\ and\ \bibinfo {author} {\bibfnamefont {S.}~\bibnamefont
  {Granick}},\ }\href@noop {} {\bibfield  {journal} {\bibinfo  {journal}
  {Langmuir}\ }\textbf {\bibinfo {volume} {16}},\ \bibinfo {pages} {8368}
  (\bibinfo {year} {2000})}\BibitemShut {NoStop}%
\bibitem [{\citenamefont {Kupchinov}\ \emph {et~al.}(1991)\citenamefont
  {Kupchinov}, \citenamefont {Rodnenkov}, \citenamefont {Ermakov},\ and\
  \citenamefont {Parkalov}}]{KUPCHINOV}%
  \BibitemOpen
  \bibfield  {author} {\bibinfo {author} {\bibfnamefont {B.}~\bibnamefont
  {Kupchinov}}, \bibinfo {author} {\bibfnamefont {V.}~\bibnamefont
  {Rodnenkov}}, \bibinfo {author} {\bibfnamefont {S.}~\bibnamefont {Ermakov}},
  \ and\ \bibinfo {author} {\bibfnamefont {V.}~\bibnamefont {Parkalov}},\
  }\href {\doibase https://doi.org/10.1016/0301-679X(91)90059-I} {\bibfield
  {journal} {\bibinfo  {journal} {Tribology International}\ }\textbf {\bibinfo
  {volume} {24}},\ \bibinfo {pages} {25 } (\bibinfo {year} {1991})}\BibitemShut
  {NoStop}%
\end{thebibliography}
%

\begin{figure}[t]
 \begin{center}
 \includegraphics{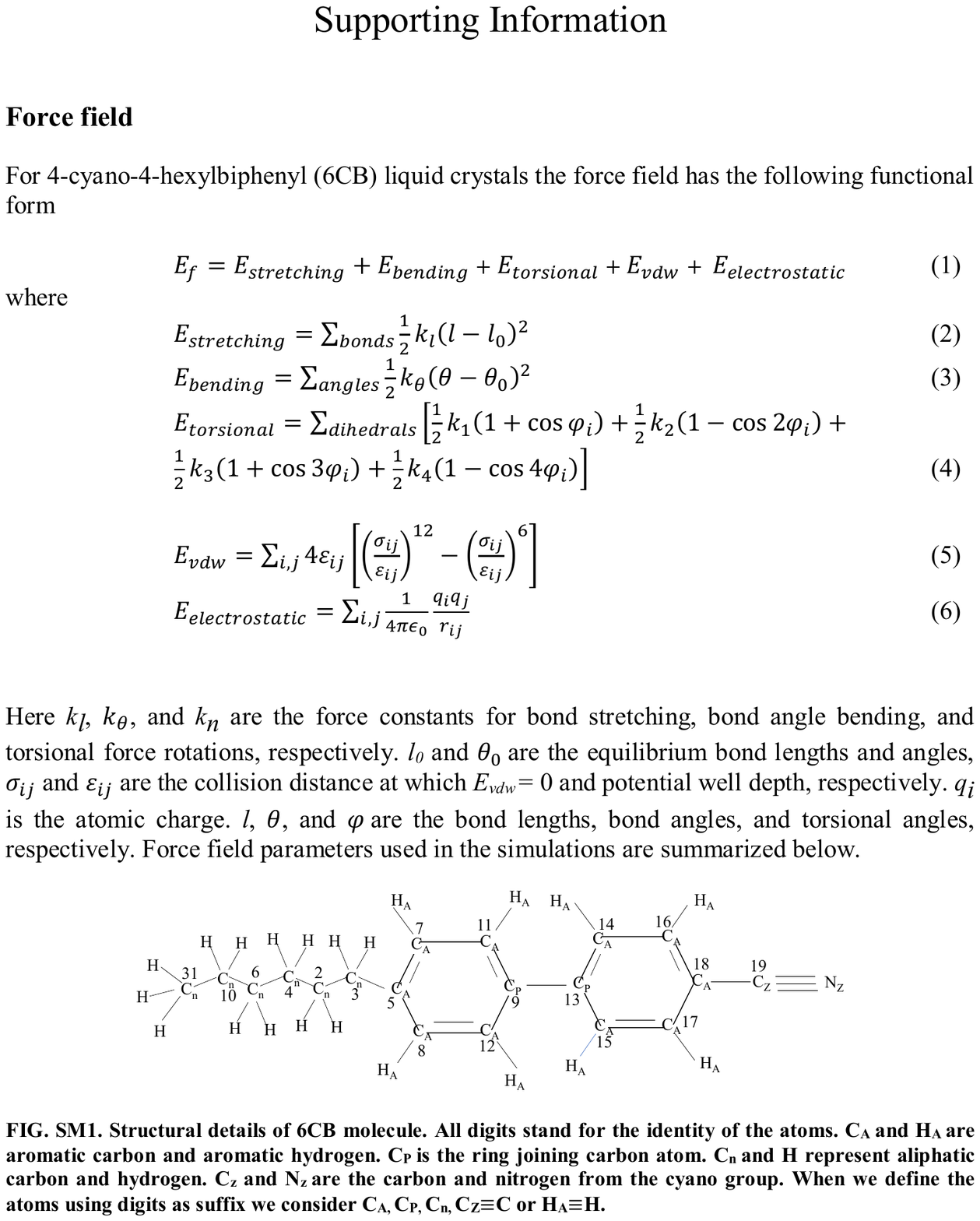}
 \end{center}
\end{figure}
\begin{figure}[t]
 \centering 
  \includegraphics{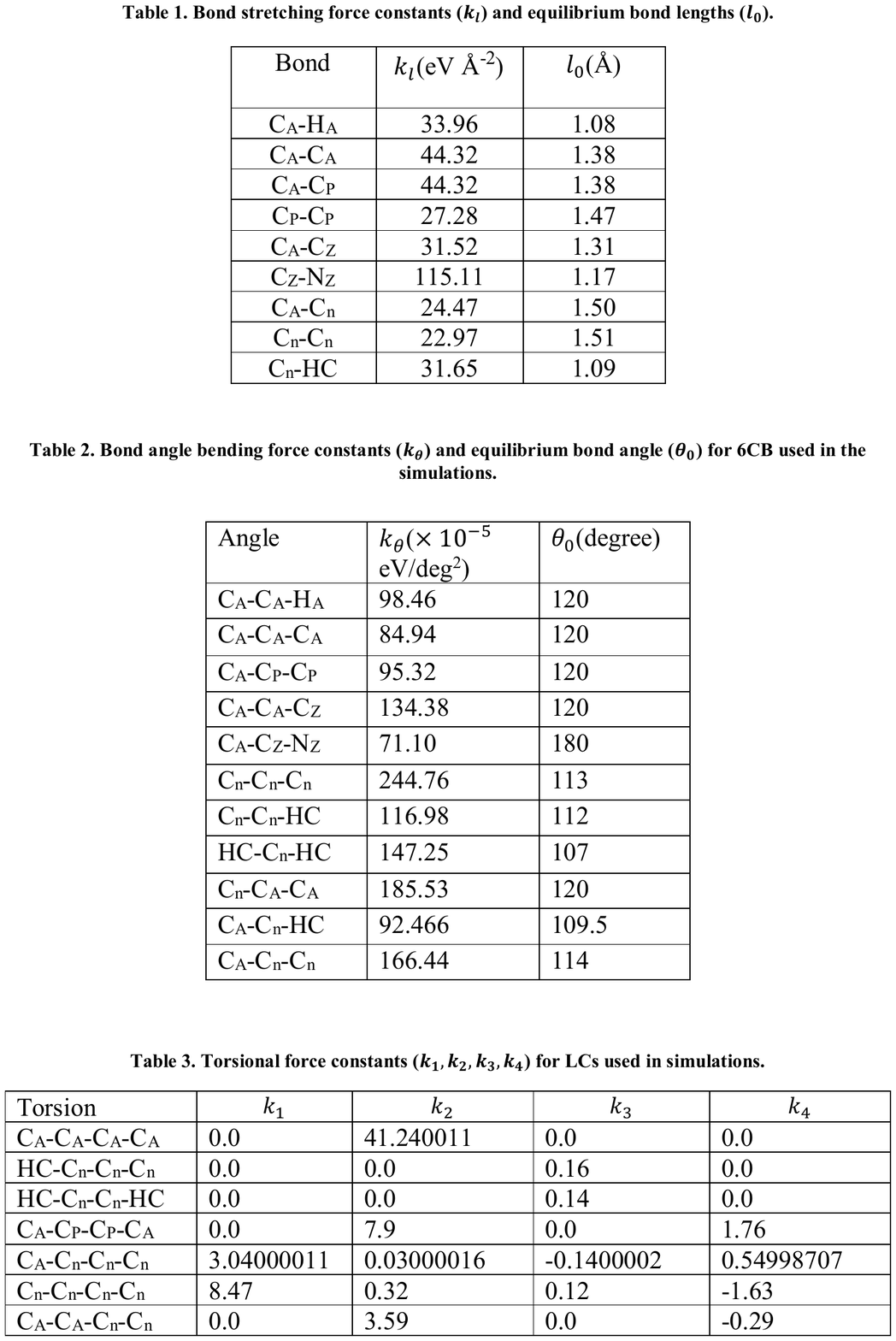}
\end{figure}
\begin{figure}[t]
 \centering 
  \includegraphics{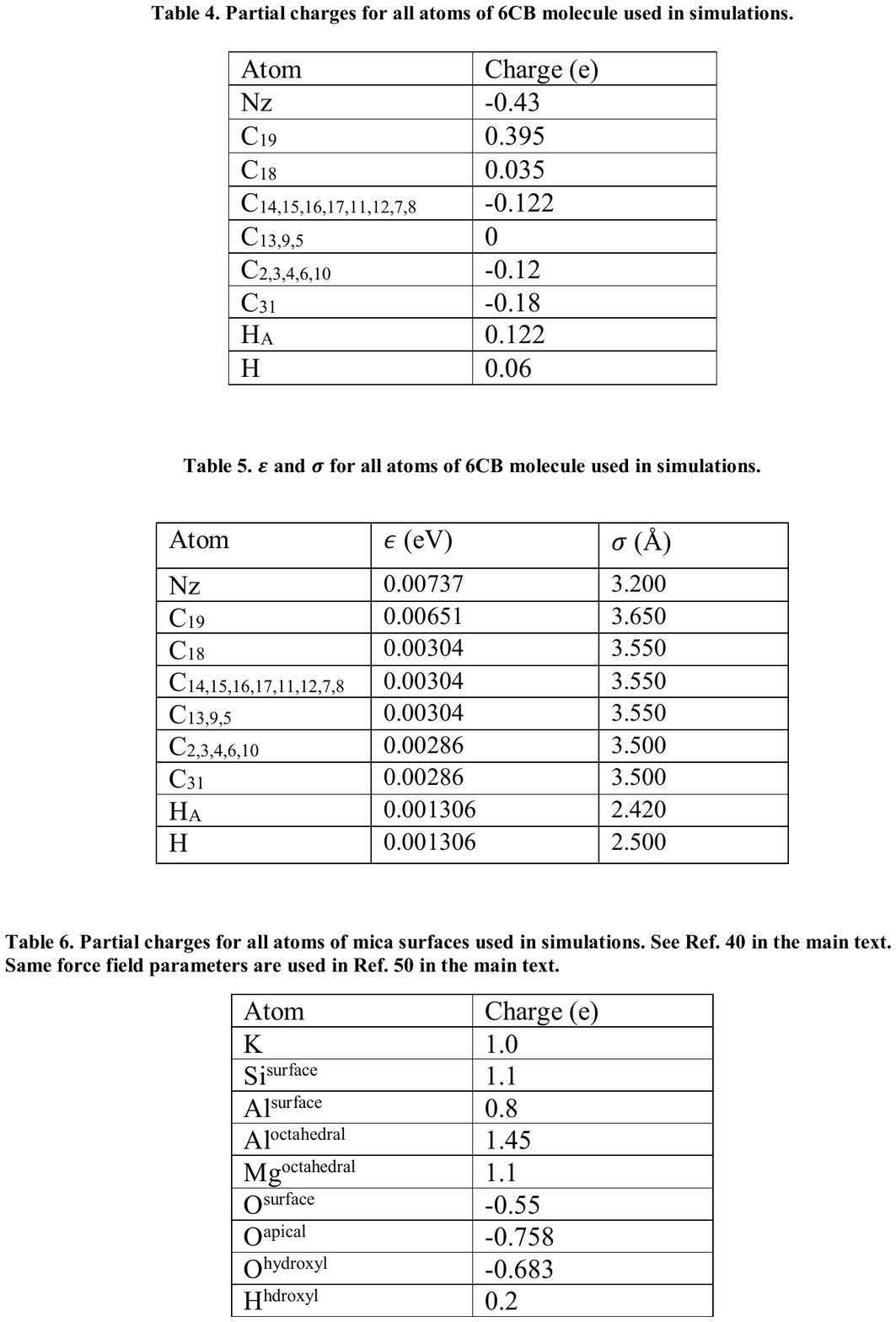}
\end{figure}
\begin{figure}[t]
 \centering 
 \includegraphics{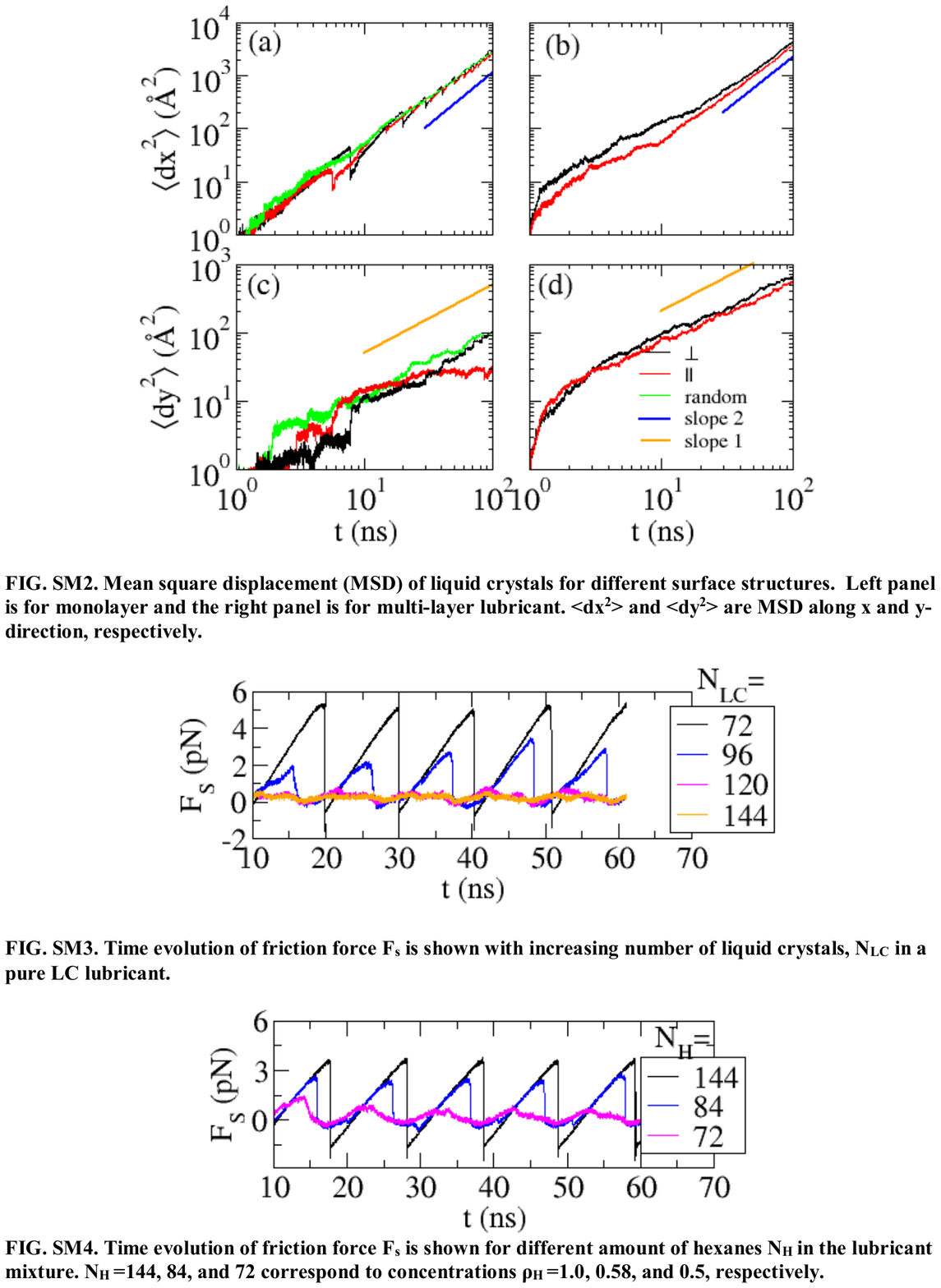}
\end{figure}
\begin{figure}[t]
 \centering 
  \includegraphics{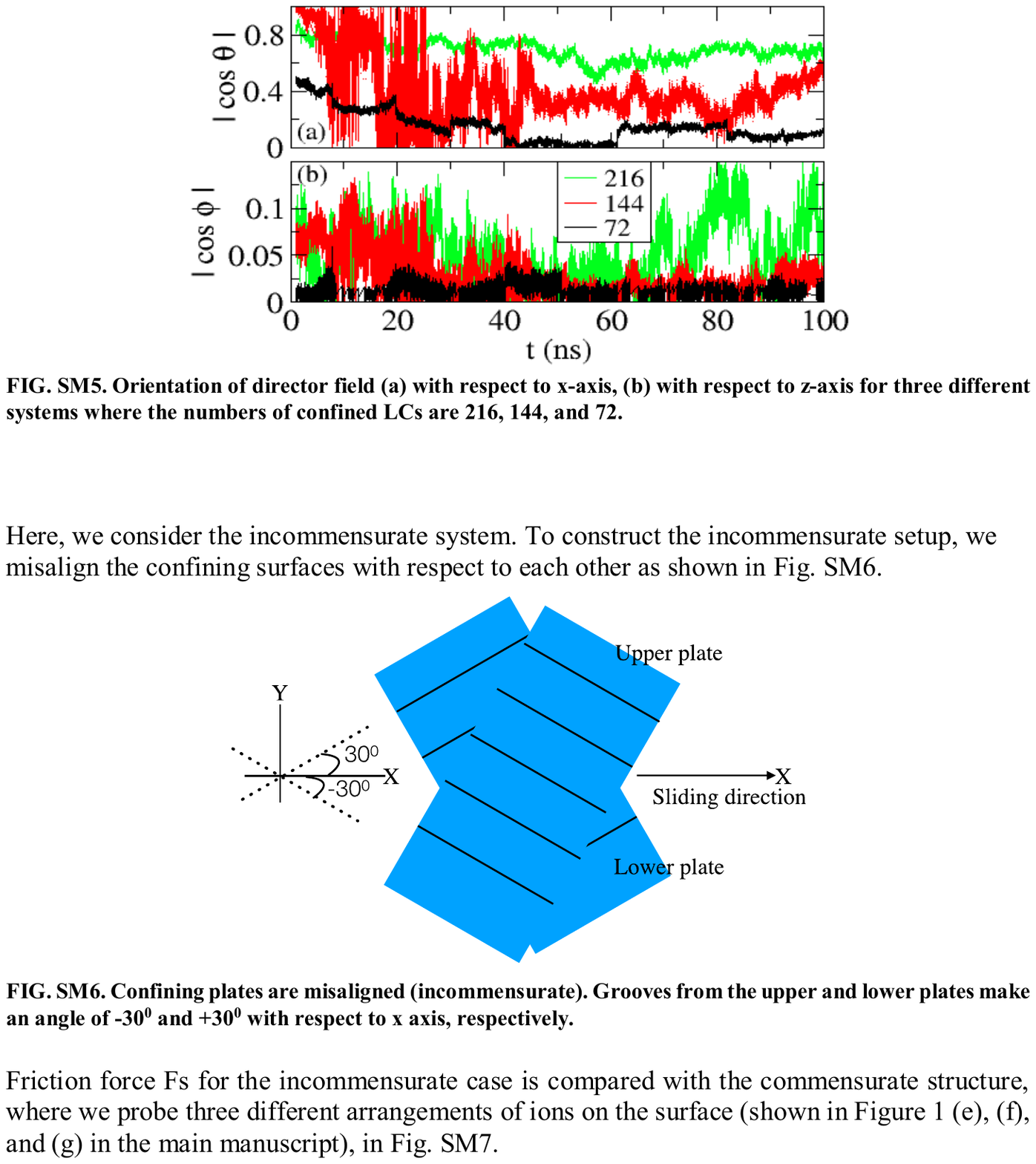}
\end{figure}
\begin{figure}[t]
 \centering 
  \includegraphics{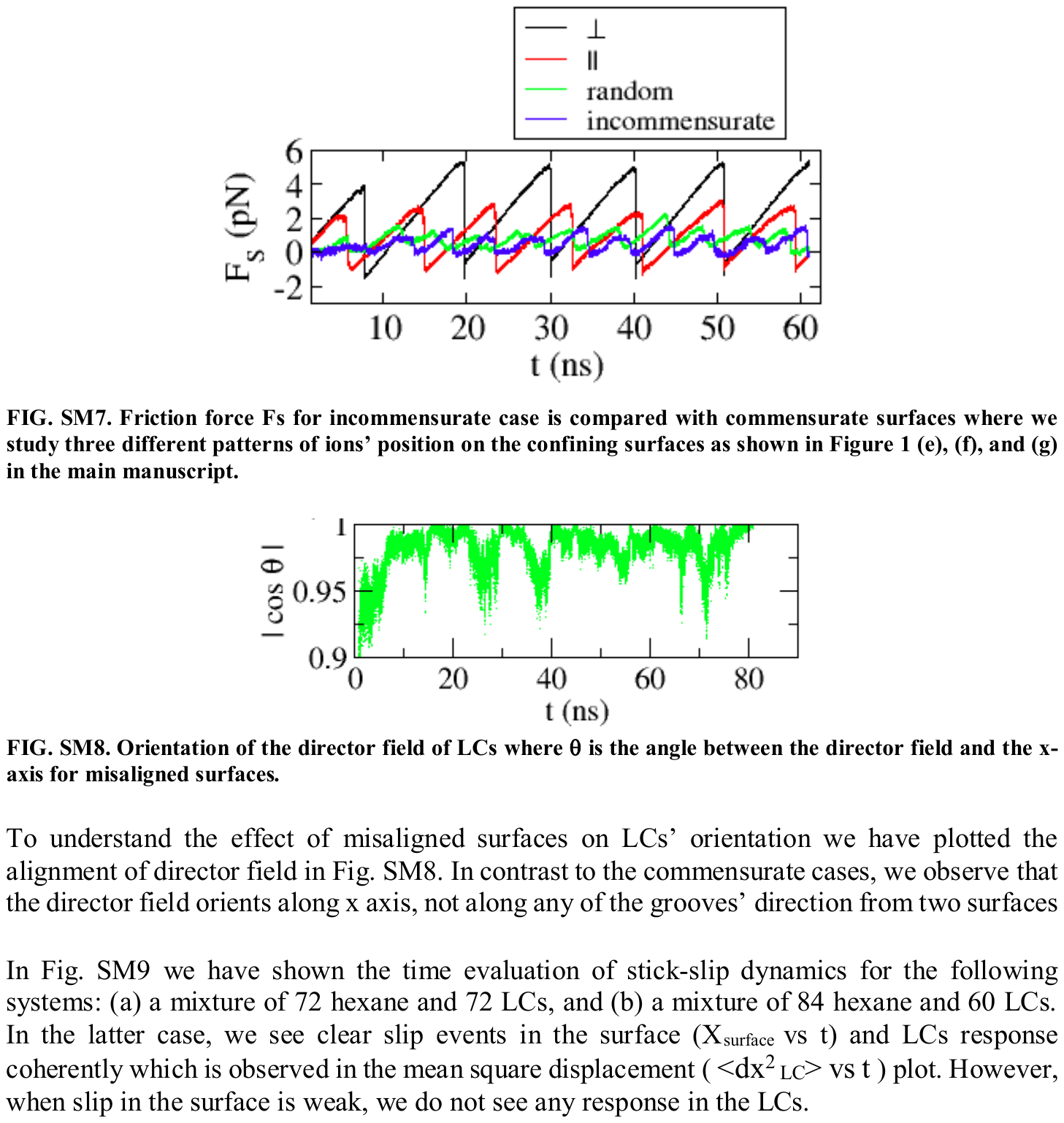}
\end{figure}
\begin{figure}[t]
 \centering 
  \includegraphics{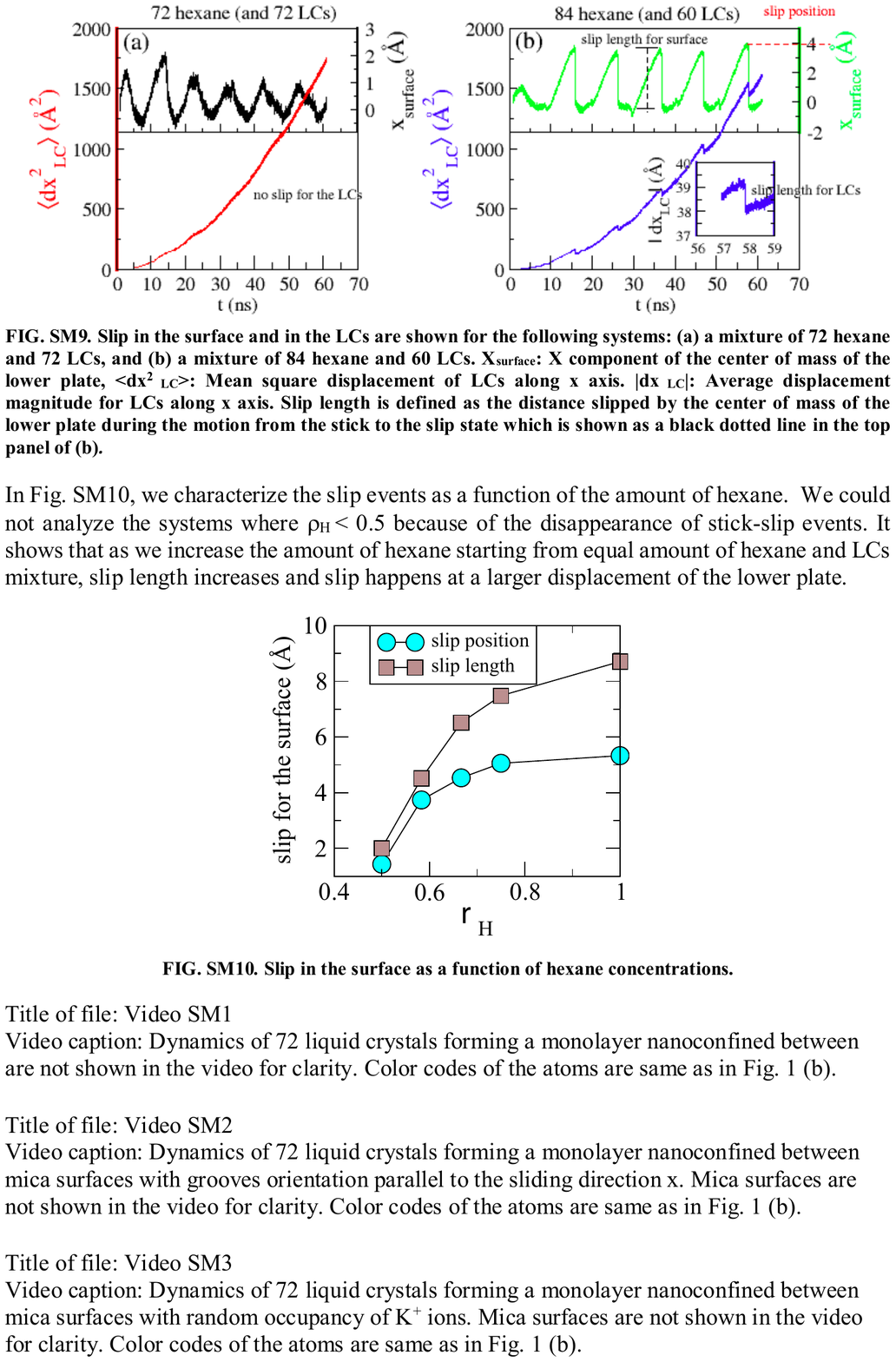}
\end{figure}
\begin{figure}[t]
 \centering 
  \includegraphics{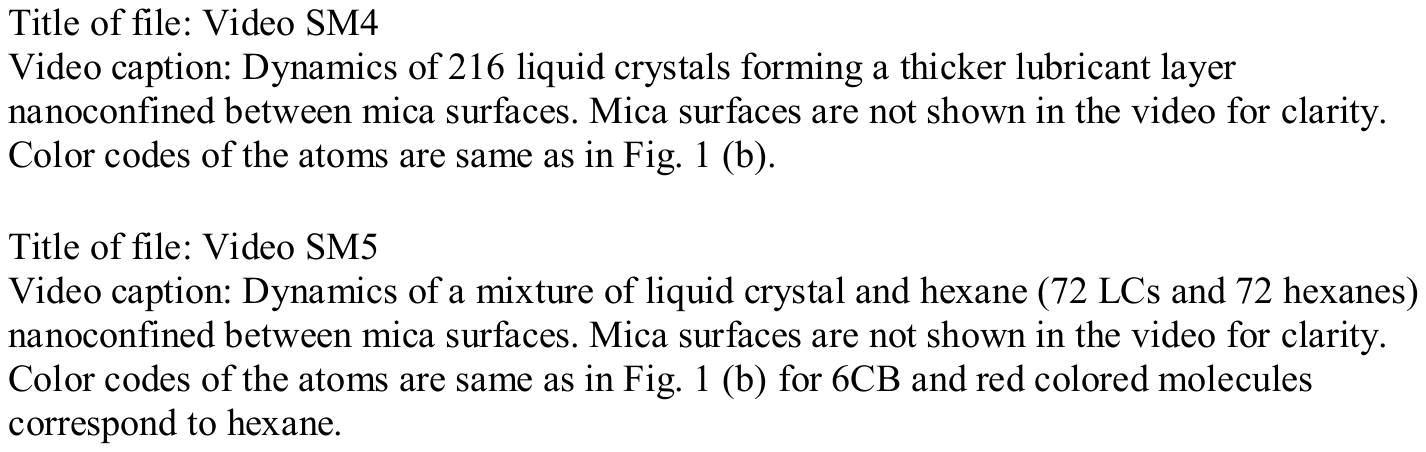}
\end{figure}

\end{document}